\documentclass[rmp,aps,nofootinbib,preprint,showpacs,amsmath,amssymb]{revtex4}
\usepackage{epsfig}
\usepackage{bm}

\begin{document}
\newcommand{\inner}{\!\cdot\!}
\newcommand{\dint}{\int\mkern-12mu\int}
\newcommand{\doint}{{\bigcirc\mkern-25mu\int\mkern-14mu\int}}
\newcommand{\cross}{\times}
\newcommand{\curl}{\nabla\times}
\newcommand{\pt}{\partial}
\newcommand{\Lk}{\hbox{\sl Lk}}
\newcommand{\Wr}{\hbox{\sl Wr}}
\newcommand{\Tw}{\hbox{\sl Tw}}
\newcommand{\Wrp}{\widetilde{\hbox{\sl Wr}}}
\newcommand{\lt}{\left}
\newcommand{\rt}{\right}
\newcommand{\perpp}{{\scriptscriptstyle\perp}}
\newcommand{\kb}{k_{\scriptscriptstyle\rm B}}
\newcommand{\kbT}{k_{\scriptscriptstyle\rm B}T}
\newcommand{\vecn}{{\bf n}}
\newcommand{\ofx}{({\bf x})}
\newcommand{\ofxp}{({\bf x}')}
\newcommand{\ofxt}{}
\newcommand{\hef}{$^4${\rm He}}
\newcommand{\het}{$^3${\rm He}}
\newcommand{\free}{\hbox{$\cal F$}}
\newcommand{\nabb}{{\bm{\nabla}}}
\newcommand{\vecR}{{\bf R}}
\newcommand{\vecX}{{\bf X}}
\newcommand{\vecT}{{\bf T}}
\newcommand{\vecN}{{\bf N}}
\newcommand{\vecB}{{\bf B}}
\newcommand{\grad}{{\bm{\nabla}}}
\newcommand{\ofss}{(s_1,s_2)}
\newcommand{\Tr}{{\rm Tr}}

\title{The Geometry of Soft Materials: A Primer}

\author{Randall D. Kamien}
\email{kamien@physics.upenn.edu}
\affiliation{
Department of Physics and
Astronomy, University of Pennsylvania, Philadelphia, PA 19104}
\date{6 March 2002}
\begin{abstract}
We present an overview of the differential geometry of curves and surfaces using
examples from soft matter as illustrations.  The presentation requires a background 
only in vector calculus and is otherwise self-contained.
\end{abstract}
\maketitle
\tableofcontents

\section{Introduction}
Though geometry is a common part of our early schooling, a rigorous and thorough education
in physics usually tries to expunge it from our thought.  Because of their predictive powers, there
is good reason to emphasize numbers and formul\ae .  Analytic geometry is the usual emphasis, while classical geometry is relegated to popular expositions.  Differential geometry is a bridge between shapes and analytic expressions and is often the appropriate language for modern physics.  Nonetheless, when necessary, geometry is often slipped in as a bitter or, at least, tasteless pill -- just enough is presented to get on with the analysis.  In these lectures I have
made an attempt to introduce the basics of differential geometry in the style of
a mathematics text: the ideas are grouped by mathematical subject as opposed to physical subject.
Nonetheless, I follow each newly developed topic with an example from
soft matter, not only to illustrate the usefulness of the mathematical structure but also
to aid the reader with a concrete example.

This is by no means a textbook and many details of the physics are left for the diligent reader to find in the references.  For technical details of many of the topics discussed here, the reader
is referred to {\sl Elements of Differential Geometry} by Millman and Parker \cite{MP}, though
any standard reference on classical differential geometry should suffice.  A word on
notation: I have tried throughout to explicitly display the functional dependence of
all the fields and functions in formulae.  However, sometimes this would make
the notation awkward and the dependencies are dropped.  In each case the context should
make clear any lack of precision.

Finally, I have tried to reduce as much as possible the use of the powerful formalism of
differential geometry.  While this formalism is useful for performing complex
calculations unambiguously, great expertise is often required to extract the geometrical
and physical meaning of these calculations.  An excellent complement to these notes are the
lecture notes by Fran\c cois David \cite{David} which present the mathematical elegance and logical compactness
of this subject.

\section{Local Theory of Curves}
\subsection{Conformations of Polymers: {\sl Motivating
a Geometrical Description of Curves}}

Random walks abound in physics.  They are the basis for
understanding diffusion, heat flow and polymers.  However,
polymers are the most interesting of the three: polymers, unlike
diffusing particles leave a ``tail'' behind them which they must
avoid.  They are described by self-avoiding random walks.  As an
introduction to the power of geometrical modelling, in this section we will
consider the behavior of stiff polymers at the shortest lengthscale
amenable to a continuum analysis.  At these scales,
polymers are not random walks at all, but should be thought of as
stiff rods.  We take this as our starting point.  When we describe
a polymer as stiff, we are ascribing to it an energy cost for
being bent.  To model this, we consider a curve $\vecR(s)$,
parameterized by its arclength $s$ and construct the
tangent vector to our curve $\vecT(s)=d\vecR(s)/ds$ at a point $s$
on the curve.  In the next section we will show that the tangent vector
is of unit length. 
If the tangent vector is constant along the curve
then it is a straight line and does not bend. Thus the energy
should depend on derivatives of the unit tangent vector
$\vecT(s)$. Indeed we call the magnitude $d\vecT(s)/ds$ the curvature of the
curve.  In the next section we will discuss the geometry of curves in greater detail.
In the
meantime, we write the energy as:
\begin{equation}\label{ecurvature}E_{\rm curv} = \frac{1}{2}A\int_0^L
\left[\frac{d\vecT(s)}{ds}\right]^2 ds,\end{equation}
where $s$ is a
parameter that measures the arclength of the curve, and $A$ is a
measure of the stiffness. To study the statistical mechanics of
the curve we write the partition function for $\vecT$:
\begin{equation}\label{epartition}Z\left(\vecT_L\right) = \int_{\vecT(0)=\hat
z}^{\vecT(L)=\vecT_L} [d\vecT] e^{-E_{\rm curv}[\vecT]/\kbT}\end{equation}
This
is the partition function for the curve which starts with its
tangent vector along the $z$-direction and ends with its tangent
vector equal to $\vecT_L$.  Note that if we can calculate
$\langle\,\vecT(s)\vecT(s')\,\rangle$, then we can integrate with
respect to $s$ and $s'$ to obtain:
\begin{equation}\label{eallofit}\left\langle\,\big(\vecR(L)-\vecR(0)\big)^2\,\right\rangle
= \int_0^L ds\int_0^L ds' \left\langle\,\frac{d\vecR(s)}{
ds}\cdot\frac{d\vecR(s')}{ds}\,\right\rangle\end{equation}
Calculating the
correlation function of the tangent vectors proves to be
straightforward.  Though there are many ways to proceed \cite{DE}, we
will choose here an analogy with quantum mechanics.  In
(\ref{epartition}) let $s=it$.  Then the partition function
becomes:
\begin{equation}\label{epartitioni}Z\left(\vecT_L,L\right) =
\int_{\vecT(0)=\hat z}^{\vecT(L)=\vecT_L} [d\vecT]
\exp\left\{{i \over\kbT} \int_0^{-iL}dt \,{A\over
2}\left({d\vecT\over dt}\right)^2\right\}\end{equation}
we recognize this as
the path integral solution to Schr\"odinger's equation for a single
particle where $\kbT$ replaces $\hbar$, $A$ replaces
the particle's mass, and $\vecT$ is the position of the particle.
Since $\vecT(t)$ lives on the unit sphere, this is just quantum
mechanics on a sphere.  The Schr\"odinger equation is:
\begin{equation}\label{eschro} -{\kbT\over i} {\partial Z\over \partial L} =
-{(\kbT)^2\over 2A}{\bf\hat L}^2Z,\end{equation} 
where $\bf\hat L$ is the
angular momentum operator.  Changing back to our original
coordinate $s$ and defining $L_p\equiv A/\kbT$ we have
\begin{equation}\label{efp}{\partial Z({\bf T},s)\over \partial s} = {1\over
2L_p}{\bf\hat L}^2 Z({\bf T},s).\end{equation}
We will soon discover that $L_p$
has an important interpretation.

Since the polymer is the same all along its length, we have
$\langle\,\vecT(s)\cdot\vecT(s')\,\rangle =
\langle\,\vecT(s-s')\cdot\vecT(0)\,\rangle$.  Using polar
coordinates and enforcing the limits of integration in (\ref{epartitioni}) so that $\vecT(0)=\hat z$, we discover
that we are interested in $\langle\,\cos\theta(S)\,\rangle$ at
$S=s-s'$:
\begin{equation}\label{ecosines}\langle\,\cos\theta(S)\,\rangle =
{\int_{-1}^1 d(\cos\theta)\cos\theta
Z(\cos\theta,S)\over\int_{-1}^1 d(\cos\theta) Z(\cos\theta,S)}\end{equation}
where we have been sure to normalize $Z$ to get a probability. The
expectation value we seek satisfies a differential equation since
$Z(\cos\theta,S)$ satisfies (\ref{efp}).  Moreover, since ${\bf\hat L}^2$
is a Hermitian operator, we have
\begin{equation}\label{edeq}
{d\langle\,\cos\theta(S)\,\rangle\over dS} =  {1\over 2L_p}
\langle\, {\bf\hat L}^2\cos\theta(S)\,\rangle = -{1\over
L_p}\langle\,\cos\theta(S)\,\rangle,\end{equation}
where we have used the fact that ${\bf\hat
L}^2 \cos\theta = -2\cos\theta$ and ${\bf\hat L}^2(1)=0$.  It follows that
\begin{equation}\label{esolut}\langle\,\cos\left[\theta(s)-\theta(s')\right]\,\rangle
= e^{-\vert s-s'\vert/L_p}.\end{equation}
Integrating this expression as in
(\ref{eallofit}), we have
\begin{equation}\label{esolofit}\left\langle\,\big(\vecR(L)-\vecR(0)\big)^2\,\right\rangle
= 2L_p\left(L-L_p + L_pe^{-L/L_p}\right).\end{equation} 
From (\ref{esolut}) and
(\ref{esolofit}) we glean the meaning of $L_p$.  The first equation
shows that the tangent vectors along the curve are uncorrelated
after a distance $L_p$.  For this reason $L_p$ is called the
persistence length \cite{dg}.  The second equation shows us that for
$L$ much longer than $L_p$, a stiff rod behaves as a random walk:
{\sl i.e.} the average square distance that is travelled is
proportional to the number of steps or length of the walk,
$R^2\propto LL_p$. For $L$ much shorter than $L_p$, we may expand
(\ref{esolofit}) to see that $R^2\propto L^2$.

The physics of stiff rods can be used to study other phenomena.  For instance,
vortices in fluids, superfluids and superconductors have a bending stiffness arising
from self-interactions.  In the past decade the physics of stiff rods has been
adapted to study the mechanical properties of DNA \cite{markosiggia,markosiggiaii}, and has been
augmented to include twisting degrees of freedom \cite{Marko,KNLO}, the
topological constraints imposed by self-avoidance \cite{selfav} and by closed loops \cite{moroz,
phil,philii},
and the effects of sequence disorder \cite{philtwo,mezard}.

\subsection{The Frenet-Serret Apparatus: {\sl DNA and Other Chiral Polymers}}

\begin{figure}\epsfxsize=2.5truein
\epsfbox{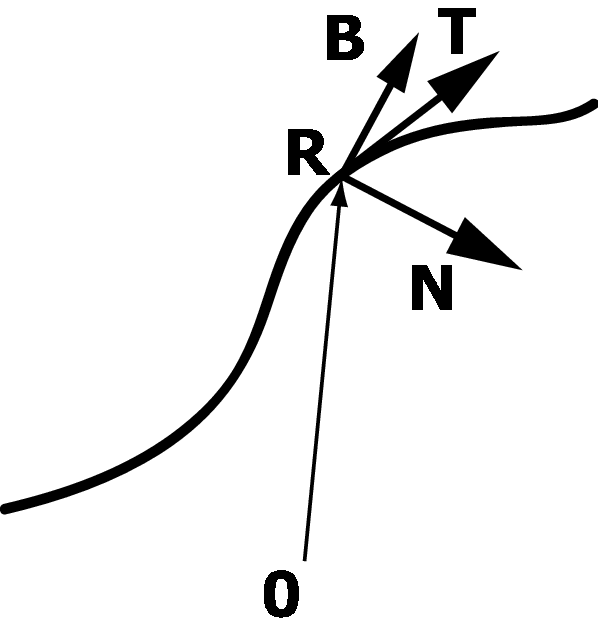}
\vskip10pt
\caption{A curve in space.  At $\bf R$ there is a tangent
vector $\vecT$, a normal vector $\vecN$ and the binormal vector $\vecB$.}
\end{figure}
\vskip10pt
In the previous section we have seen that a simple geometrical description of 
polymers leads to a precise description of their conformational behavior in a variety
of regimes.  It is now time to discuss more carefully the geometry that went into making
the expression for energy in (\ref{ecurvature}).
A curve in three dimensions
is a vector-valued function $\vecR(s)=[x(s),y(s),z(s)]$
that depends on $s$, a parameter that runs along the curve.
Though we may choose to label points along the curve as we wish,
it is usually most convenient to let $s$ be the arclength along
the curve.  We will see how this simplifies our equations shortly.
If the curve is $L$ long, then $s$ runs from $0$ to $L$.  The
first thing to do is to construct the unit tangent vector to the
curve: as we have already asserted, the magnitude of the rate of change of the unit
tangent vector is the curvature $\kappa$ of the curve.  We have
\begin{equation}\label{etangent}\vecT(s) = {\vecR'(s)\over\vert\vert
\vecR'(s)\vert\vert},\end{equation}
where ${\bf X}'(s)$ denotes the derivative
of $X(s)$ with respect to $s$, and $\vert\vert {\bf X}\vert\vert$
is the length of the vector $\bf X$.   We have divided by the
length of $\vecR'(s)$ to make $\vecT(s)$ a unit vector.  However,
if $s$ measures the arclength along the curve then
$\vert\vert\vecR'(s)\vert\vert=1$!  To see this, recall that the
length of a curve $\vecR(t)$ from $t=0$ to $t=t_f$ can be found by
adding (integrating) the length of pieces of the curve together,
each $\sqrt{d\vecR(t)\cdot d\vecR(t)}$ long:
\begin{equation}\label{elength}L(t_f) =
\int_0^{t_f} \sqrt{d\vecR(t)\cdot d\vecR(t)} = \int_0^{t_f}
dt\,\sqrt{{d\vecR(t)\over dt}\cdot{d\vecR(t)\over dt}}.\end{equation}
If we
choose $t=s$ to be the arclength then we have
\begin{equation}\label{especlength}L =
\int_0^L ds \vert\vert\vecR'(s)\vert\vert,\end{equation}
where the upper limit
is the same as the length of the curve.  Differentiating both
sides of (\ref{especlength}), we see that
$\vert\vert\vecR'(s)\vert\vert=1$.

Having constructed the unit tangent vector, we may now take
its derivatives to obtain the curvature.  Since the derivative
of a vector is another vector, we write
\begin{equation}\label{enormal}\vecT'(s) =\kappa(s)\vecN(s)\end{equation}
where $\kappa(s)$ is the curvature at $s$ and
$\vecN(s)$ is a new vector, {\sl the unit normal vector}, which is
also unit length, so that
$\vert\vert\vecT'(s)\vert\vert=\vert\kappa(s)\vert$.   It is
convention to choose $\kappa(s)$ to be always positive -- the sign
can alway be absorbed in the direction of $\vecN(s)$. Note that
$\vecT(s)\cdot\vecN(s)=0$ since the derivative of {\sl any} unit
vector is perpendicular to itself; if $\vert\vert{\bf
X}(s)\vert\vert=1$ then
\begin{eqnarray}\label{eperpendicular}{d\over
ds}\left[{\bf X}(s)\cdot{\bf X}(s)\right] &= &{d\over ds}[1]\nonumber\\
2{\bf X}(s)\cdot{\bf X}'(s) &=&0. \end{eqnarray}
Of course, as we move along
$s$, the normal vector changes direction as well.  Changes in the
direction of the normal vector can come from two contributions:
the normal can change by rotating towards or away from the tangent
vector (of course the pair rotate together to stay
perpendicular).  It can also change by rotating {\sl around} the
tangent vector.  The former case corresponds to the curve staying
in the same, flat plane, while the second corresponds to rotations
of the plane in which the curve lies at $s$.  This plane is
known as the {\sl osculating} plane, from the Greek word for
kissing.  Moreover, since $\vecN'(s)$ is perpendicular to
$\vecN(s)$, we must introduce a third unit vector to account for
changes in the osculating plane.  We choose
$\vecB(s)=\vecT(s)\times\vecN(s)$ where $\times$ denotes the cross
product, as shown in Figure 1. This vector, {\sl the binormal vector}, is a unit vector
perpendicular to both $\vecT(s)$ and $\vecN(s)$.  We then have
\begin{equation}\label{etorsion}\vecN'(s) = \alpha(s)\vecT(s) + \tau(s)\vecB(s)\end{equation}
where $\alpha(s)$ is some function of $s$ and $\tau(s)$ is called
the {\sl torsion} of the curve.  It is a measure of the rate of
change of the osculating plane.  Why don't we give $\alpha(s)$ a
name?  Because we note that by differentiating the relation
$\vecT(s)\cdot\vecN(s)=0$ we get:
\begin{equation}\label{ealphakappa}0=\vecT'(s)\cdot\vecN(s) + \vecT(s)\cdot\vecN'(s)
= \kappa(s) +\alpha(s),\end{equation}
so $\alpha(s)=-\kappa(s)$ and (\ref{etorsion})
becomes:
\begin{equation}\label{etorsionii}\vecN'(s)=-\kappa(s)\vecT(s) +
\tau(s)\vecB(s).\end{equation}

Finally, we may calculate $\vecB'(s)$ to complete our analysis of
the curve.  We have
\begin{eqnarray}\label{ebinormal}\vecB'(s) &=&
\vecT'(s)\times\vecN(s) + \vecT(s)\times\vecN'(s)\nonumber\\
&=&
\kappa(s)\vecN(s)\times\vecN(s) -\kappa(s)\vecT(s)\times\vecT(s)
+\tau(s)\vecT(s)\times\vecB(s)\cr &= &-\tau(s)\vecN(s)
\end{eqnarray}
where
the last line follows from the rule ${\bf a}\times\left({\bf
b}\times{\bf c}\right) = {\bf b}\left({\bf a}\cdot{\bf c}\right)
-{\bf c}\left({\bf a}\cdot{\bf b}\right)$. Putting (\ref{enormal}),
(\ref{etorsionii}), and (\ref{ebinormal}) together we have the Frenet-Serret
equations for a curve in three-dimensions:
\begin{equation}\label{efs}{d\over
ds}\left[\begin{matrix}\vecT(s)\\ \vecN(s)\\ \vecB(s)\end{matrix}\right] 
=\left[
\begin{matrix}
0 &\kappa(s)&0\\ -\kappa(s)&0&\tau(s)\\
0&-\tau(s)&0\end{matrix}\right]
\left[\begin{matrix}\vecT(s)\\ \vecN(s)\\ \vecB(s)\end{matrix}\right]\end{equation}
This shows that $\kappa(s)$ is the rate of rotation of $\vecT(s)$ about $\vecB(s)$ and similarly,
$\tau(s)$ is the rate of rotation of $\vecN(s)$ about $\vecT(s)$.
Written as one matrix equation, the Frenet-Serret formula tell us
something very important:  given a curvature $\kappa(s)$ and a
torsion $\tau(s)$, we can reconstruct our entire curve up to a
translation (since we can change the origin) and a rotation (since
we can rotate the initial orthonormal triad
$\{\vecT(0),\vecN(0),\vecB(0)\}$.  Once we have set the location
and orientation of that triad, however, the entire curve is
determined by only two parameters, not three as one might have thought.

There is
a difficulty with the Frenet-Serret frame: when the curvature vanishes, $\vecN(s)$ is
not well-defined and therefore $\vecB(s)$ and, more importantly, the torsion are
not defined either.  Thus curves that have straight segments are problematic
from the point of view of the Frenet-Serret frame.  Moreover, if we consider a helix
\begin{equation}\label{ehelix}\vecR(s) = \left[r\cos\left({qs\over\sqrt{(qr)^2+1}}\right),r\sin\left({qs\over\sqrt{(qr)^2+1}}\right), {s\over\sqrt{(qr)^2+1}}\right]\end{equation}
then the curvature and torsion are constant:
\begin{eqnarray}\label{etorcurv}
\kappa(s)&=& {q^2r\over 1+(qr)^2}\nonumber\\
\tau(s)&=&{q\over 1 +(qr)^2}. \end{eqnarray}
As $r\rightarrow 0$ we approach a straight line.  However, though the curvature
vanishes in this limit, the torsion does not!  This is a problematic feature of the Frenet-Serret frame -- since the torsion is the magnitude of the derivative of $\vecN(s)$, it is only
a meaningful quantity when $\vecN(s)=\vecT'(s)/\kappa(s)$ is unambiguously defined, and
this requires that $\kappa(s)\ne 0$.
In \S{\bf III}B, we will offer a different frame that does not suffer from this problem.

Knowing that the there are only two parameters needed to describe a space curve, we can
now augment (\ref{ecurvature}) to include other effects.  One interesting effect \cite{HKL} is the behavior
of {\sl chiral} polymers.  While the curvature does not distinguish between left and right handed helices, we
can see from (\ref{etorcurv}) that the torsion is sensitive to the sign of $q$.  Thus we might
add terms to (\ref{ecurvature})  to favor a particular chirality for the stiff polymer.  We might be tempted
to write
\begin{equation}\label{eaugmentcurv}E = E_{\rm curv} + E^* = \int_0^L ds\left\{ {A\over 2}\kappa^2(s) + {B\over 2}\left[\tau(s)-\tau_0\right]^2\right\}.\end{equation}
Though this energy appears acceptable, and favors an average torsion $\tau=\tau_0$, it only accentuates the
ambiguity of $\tau$ when $\kappa=0$.  Moreover, since (\ref{ecurvature}) is a functional of $\vecR(s)$, we should
construct the new energy only in terms of $\vecR(s)$ and its derivatives.  In addition
to $[\vecR''(s)]^2=[\vecT'(s)]^2$, we can also construct the term \cite{KNLO,Marko}
\begin{equation}\label{echiral}E^* = -{\alpha\over 2}\int_0^L ds\, \vecT(s)\cdot\left[\vecT'(s)\cross\vecT''(s)\right].\end{equation}
From the Frenet-Serret formula, we then have:
\begin{equation}\label{eaugment}E= \int_0^L ds\,\left\{{A\over 2}\kappa^2(s) +{C\over 4}\kappa^4(s)- {\alpha\over 2}\kappa^2(s)\tau(s) + {\beta\over 2}\kappa^2(s)\tau^2(s)\right\},\end{equation}
where $A$, $C$, $\alpha$ and $\beta$ are all positive.  Note that in the special case of constant curvature and torsion, $\left[\vecT(s)\cross\vecT''(s)\right]^2=\kappa^2\tau^2$.  
We have added this term and the extra quartic term to $E$ for reasons that will become clear in the following.  This form does not suffer from the torsion ambiguity: when $\kappa(s)=0$ the torsion
drops out of (\ref{eaugment}). One can minimize this energy for 
the helix (\ref{ehelix}), which has constant curvature
and torsion, to find the ground
state conformation of a chiral polymer.  For appropriate parameter values, the tendency
for torsion can overcome the tendency to be straight and both $\kappa$ and $\tau$ will
be nonvanishing.  

\section{Global Theory of Curves}

\subsection{Fenchel's Theorem: {\sl Energetic Bounds on Closed Curves and Knots}}
\begin{figure}\epsfxsize=2.5truein\epsfbox{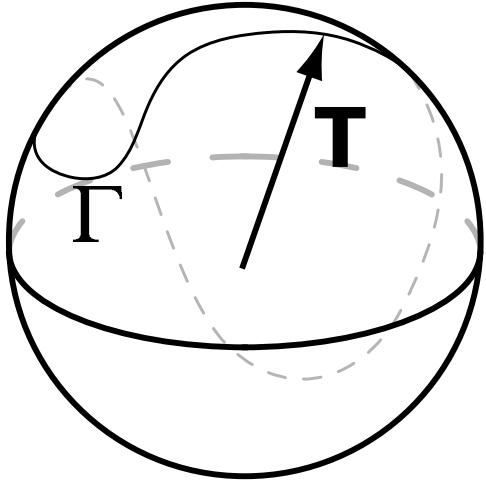}
\vskip10pt
\caption{The tangent spherical map.  The tangent vector of the curve $\vecT$ traces out a curve $\Gamma$ on the unit sphere.  The length of $\Gamma$ is the integrated
curvature along the original curve.}
\end{figure}
\vskip10pt

When we consider the energetics of
closed curves, it is clear that the curvature cannot vanish
everywhere (or the curve won't close on itself).  We can actually
establish a lower bound on the curvature energy by appealing to a
theorem on the total curvature of the curve.  Not only will this
be useful when considering closed polymer loops, the technology of
the proof will help us later on when we consider the geometry of
surfaces.

Fenchel's theorem states that the total curvature of a closed
curve is at least $2\pi$.  This is not unreasonable: a circle of
radius $R$ has curvature $1/R$ and so the total curvature is $2\pi
R/R=2\pi$.  In general, other closed curves only have {\sl more}
curvature (in fact, only planar convex curves have an integrated curvature
of exactly $2\pi$).  The statement of the theorem is:
\begin{equation}\label{efenchel}\oint_0^L \kappa(s) ds \ge 2\pi\end{equation}
for any closed
curve, where, as usual, $s$ is the arclength of the curve.  Before
we prove this fact, note that via the Cauchy-Schwarz inequality we
have
\begin{equation}\label{ecs}\oint_0^L (1)\kappa(s) ds \le \sqrt{\oint_0^L
\kappa^2(s)ds}\sqrt{\oint_0^L 1^2 ds}.\end{equation} 
Squaring both sides and
using (\ref{efenchel}) we find that the curvature energy (\ref{ecurvature})
satisfies
\begin{equation}\label{ebound}E_{\rm curv}=\frac{1}{2}\kbT L_p\oint_0^L \kappa^2(s)ds
\ge 2\pi^2\kbT{L_p\over L}.\end{equation}

To prove Fenchel's theorem we introduce the tangent spherical
image \cite{MP} of the curve:  we take the tail of the unit tangent vector $\vecT(s)$ at
$\vecR(s)$ and map it to the center of the unit
sphere.  As shown in Figure 2, the tip of the tangent vector then traces out a curve
$\Gamma$ on the surface of the unit sphere as $s$ goes from $0$ to
$L$. The differential of length of the curve on the tangent sphere
is $\vert\vert\vecT(s)'\vert\vert ds$, in analogy with
(\ref{especlength}) . But the magnitude of $\vecT'$ is just the
curvature, so we find that the length of $\Gamma$ on the unit
sphere is
\begin{equation}\label{elenonsph}\ell = \int_0^L \kappa(s)ds,\end{equation}
the total curvature of our original curve $\vecR(s)$. Now we note
that by definition of $\vecT(s)$,
\begin{equation}\label{espace}\vecR(s)-\vecR(0) = \int_0^s \vecT(s')ds',\end{equation}
so if the curve is closed, $\vecR(L)=\vecR(0)$, and the left-hand
side of (\ref{espace}) is $\bf 0$.  This is really three equations, one
for each component of $\vecT$.  It tells us that the curve
$\Gamma$ on the tangent sphere can never be in only one
hemisphere. In order for the curve to close, it must turn around.  But if the tangent spherical map were in
one hemisphere, we could take it to be the upper
hemisphere where $T_z(s)>0$.  Yet if $T_z(s)$ is always positive
then its integral cannot vanish! Therefore, given any hemisphere
on the tangent sphere, the curve must be in it and its complement.  
But this means that $\Gamma$ must be {\sl at least} the length of
the equator, $2\pi$.  We have thus proven (\ref{efenchel}). The value of
this result is not just the bound on the energetics of a closed
polymer.  It has introduced us to the tangent map.  As we will see
in the following it is very useful to take vectors off of curves
and surfaces and translate them to the center of the unit sphere.
This was our first taste of this procedure.

In closing this section we mention the Fary-Milnor theorem which
pertains to the integrated curvature of a non-self-intersecting
closed knot. Not surprisingly, this theorem states that a knot has
to go around at least twice:
\begin{equation}\label{efarymilnor}\oint_0^L\kappa(s)ds\ge 4\pi.\end{equation} 
Using the same
reasoning that led to (\ref{ebound}), we have
\begin{equation}\label{eknotbound}\kbT L_p\oint_0^L \kappa^2(s)ds\ge
(4\pi)^2\kbT{L_p\over L}\end{equation}
for a knotted closed curve.

\subsection{The Mermin-Ho Relation: {\sl Basis Vectors to the Rescue} }

Up until this point, we have been able to study the geometry of
lines without the introduction of a frame:  that is, all our results
rely on the original tangent vector $\vecT(s)$ and its derivatives -- at
no point were we required to choose a basis or coordinate system
to calculate any quantities.  Unfortunately, we cannot continue our
discussion of lines and we {\sl certainly} cannot discuss surfaces without
introducing some more paraphernalia.  We need to introduce a
set of spatially varying basis vectors in order to define
quantities in addition to the tangent vector, the curvature and the torsion.  We
need to make sure, however, that all of our {\sl physical} quantities do
not depend on our arbitrary choice.  In this section we will
derive and explain the Mermin-Ho relation \cite{MH},
originally derived
in the context of superfluid $^3$He-A, a phase characterized by an
order parameter with two directions, a ``long'' direction and a
``short'' direction perpendicular to it.  Though this section
is the most technical, it is also straightforward.  Fortunately, once
we have established this result we will be able to make use of it again and again in the following.

Often we have a vector ${\bf n}\ofx$ defined everywhere on a surface or in space.  It could
be the director of a liquid crystal, the normal to a surface or some other field of interest.  We
now are interested in a vector ${\bf N}\ofx$ which is always perpendicular
to ${\bf n}\ofx$.  This vector might point to the nearest neighbor or in some special
direction on the surface.  To define this vector, we introduce two new unit vectors, ${\bf e}_1\ofx$
and ${\bf e}_2\ofx$, which are defined to be perpendicular to ${\bf n}\ofx$ and each other so
that $\{{\bf e}_1\ofx,{\bf e}_2\ofx,{\bf n}\ofx\}$ is an orthonormal triad and ${\bf e}_1\ofx\cross{\bf e}_2\ofx ={\bf n}\ofx$.
A vector which is defined relative to the spatially
varying plane defined by ${\bf e}_1\ofx$ and ${\bf e}_2\ofx$ will
change not only because its true direction changes (relative to a
fixed triad) but also because the basis vectors change.  If ${\bf N}\ofx$ is a unit vector then
${\bf N}\ofx = \cos[\theta\ofx]{\bf e}_1\ofx
+\sin[\theta\ofx]{\bf e}_2\ofx$ is always perpendicular to $\bf n$.
Its derivatives in the $i$ direction are vectors as well.   In the plane normal to $\bf n$ their
components are:
\begin{eqnarray}\label{econstantinspacea}
{\bf e}_1\ofx\cdot\partial_i{\bf
N}\ofx &=&-\sin\theta\ofx \left[\partial_i\theta\ofx - {\bf
e_1}\ofx\cdot\partial_i{\bf e}_2\ofx\right]\\ 
{\bf e}_2\ofx\cdot\partial_i{\bf N}\ofx &=&\cos\theta\ofx
\left[\partial_i\theta\ofx + {\bf e}_2\ofx\cdot\partial_i{\bf
e_1}\ofx\right]\nonumber\\ \label{econstantinspaceb}
&=&\cos\theta\ofx\left[\partial_i\theta\ofx -
{\bf e}_1\ofx\cdot\partial_i{\bf e}_2\ofx\right] \end{eqnarray}
where we
have used the fact that ${\bf e}_1\ofx\cdot{\bf e}_2\ofx=0$ in the
final equality. Notice that the derivatives of ${\bf N}\ofx$ depend on both
gradients of $\theta\ofx$ as well as derivatives of the spatially varying basis vectors ${\bf e}_i\ofx$.  Since the basis vectors were chosen arbitrarily, one might be concerned that the derivatives of ${\bf N}\ofx$ are poorly defined.  However, there is a concomitant change in $\theta\ofx$ whenever the basis vectors change so that the gradients of ${\bf N}\ofx$ are well defined.  The problem
is with gradients of $\theta\ofx$.

To disentangle the arbitrary dependence on basis vectors, we
start by considering a vector ${\bf N}_\circ\ofx$ which is constant in the instantaneous plane perpendicular to ${\bf n}\ofx$ so that (\ref{econstantinspacea}) and (\ref{econstantinspaceb}) both vanish.  For ${\bf N}_\circ\ofx$ to be constant in
the plane perpendicular to ${\bf n}\ofx$, $\theta\ofx$ must be equal to some
$\theta_\circ\ofx$ with
$\grad\theta_\circ\ofx={\bf e}_1\ofx\cdot\nabb{\bf e}_2\ofx\equiv{\bm{\Omega}}\ofx$,
where the last equality defines a new vector field ${\bm{\Omega}}\ofx={\bf e}_1\ofx\cdot\nabb{\bf e}_2\ofx$ called the {\sl spin connection}.  Note that we can only solve this equation
for $\theta_\circ\ofx$ if $\nabb\cross\bm{\Omega}=0$. 
If we now consider an arbitary vector perpendicular to ${\bf n}\ofx$, ${\bf N}\ofx$, we should focus not on gradients of $\theta\ofx$, but rather on $\left[\theta\ofx-\theta_\circ\ofx\right]$, which is a measure of how much $\theta\ofx$ deviates from its ``constant'' value.  But $\nabb\theta_\circ\ofx = {\bm{\Omega}}\ofx$, so $\nabb\left[\theta\ofx-\theta_\circ\ofx\right]=\nabb\theta\ofx - {\bm{\Omega}}\ofx$.  Thus we see that by subtracting ${\bm{\Omega}}\ofx$ from $\nabb\theta\ofx$, we remove that part
of $\theta\ofx$ that is induced by a spatially varying basis.  Moreover, even if
$\nabb\cross\bm{\Omega}$ is nonvanishing, we can generalize this discussion to form
the combination
${\bf D}\theta\ofx\equiv\nabb\theta\ofx - {\bm{\Omega}}\ofx$, the {\sl covariant
derivative}.  It measures the true changes in $\theta\ofx$, relative to $\theta_\circ\ofx$.
If $\left[\theta\ofx-\theta_\circ\ofx\right]$ is a smooth field then $\nabb\cross\nabb\left[\theta\ofx
-\theta_\circ\ofx\right]=0$, or
\begin{equation}\label{ecurlofboth}
\nabb\cross\nabb\theta\ofx = \nabb\cross{\bm{\Omega}}\ofx.\end{equation}
We might be tempted to set the left hand side equal to zero, since usually the curl
of a gradient vanishes.  However, as we are about to show, the curl of ${\bm{\Omega}}\ofx$
is not always zero!  This means that there must be some sort of singularities or {\sl defects}
in $\theta\ofx$ (and $\theta_\circ\ofx$ -- the difference $\theta\ofx-\theta_\circ\ofx$ is smooth).  We will talk more about defects in \S{\bf V}b.

We have for the $i^{\rm th}$ component of the curl:
\begin{eqnarray}\label{ecurlofomega}
\left[\nabb\cross{\bm{\Omega}}\ofx\right]_i &=&
\epsilon_{ijk}\partial_j\left[e_1^\alpha\ofx\partial_ke_2^\alpha\ofx\right]\nonumber \\
&=& \epsilon_{ijk}\left[\partial_je_1^\alpha\ofx\right]\left[\partial_ke_2^\alpha\ofx\right]
+ e_1^\alpha\ofx\epsilon_{ijk}\partial_j\partial_ke_2^\alpha\ofx\nonumber\\
&=& \epsilon_{ijk}\left[\partial_je_1^\alpha\ofx\right] \left[\partial_ke_2^\alpha\ofx\right] + 0
\end{eqnarray}
where the last term vanishes due to the antisymmetry of $\epsilon_{ijk}$ and we have
used indices to make the calculation unambiguous\footnote{The only tensor on which we will rely is the {\sl antisymmetric tensor} $\epsilon_{ijk}$.  It is defined by
\[\epsilon_{ijk} = \left\{\begin{matrix}
\hfill+1&&\mbox{if $ijk$ is an even permutation of $123$}\hfill\\
\hfill-1&&\mbox{if $ijk$ is an odd permutation of $123$}\hfill\\
\hfill 0&&\mbox{if any two of $i$,$j$ or $k$ are the the same}\hfill\\ \end{matrix}\right.\].}.  We have used
the Einstein summation convention of dropping summation signs for repeated indices.  Unless
otherwise indicated, an index should be summed over if it appears twice in any formula.  Both the Greek and Roman indices run from $1$ to $3$.  Now consider
the object $\partial_je_1^\alpha\ofx$.  Since ${\bf e}_1\ofx$ is a unit vector, its derivative is perpendicular to it.  Therefore, we can write $\partial_je_1^\alpha\ofx$ in terms of the basis
vectors ${n}^\alpha\ofx$ and $e_2^\alpha\ofx$:
\begin{equation}\label{edjeone}\partial_je_1^\alpha\ofx  =
A_j\ofx n^\alpha\ofx  + B_j\ofx e_2^\alpha\ofx.\end{equation}
Similarly, we can do the same for $\partial_ke_2^\alpha\ofx$:
\begin{equation}\label{edketwo}\partial_ke_2^\alpha\ofx = C_k\ofx n^\alpha\ofx + D_k\ofx e_1^\alpha\ofx.\end{equation}
Putting these expressions into (\ref{ecurlofomega}) and using the orthogonality
of our triad, we find:
\begin{equation}\label{ecurlofomegatwo}\left[\nabb\cross{\bm{\Omega}}\ofx\right]_i
= \epsilon_{ijk} A_j\ofx C_k\ofx.\end{equation}
By taking the dot product of equation (\ref{edjeone}) with ${\bf n}\ofx$, we find
$A_j\ofx = n^\beta\ofx\partial_j e_1^\beta\ofx = - e_1^\beta\ofx\partial_j n^\beta\ofx$.
Similarly, $C_k\ofx = -e_2^\gamma\ofx\partial_k n^\gamma\ofx$ and so
\begin{equation}\label{ecurlofomegathree}\left[\nabb\cross{\bm{\Omega}}\ofx\right]_i =
e_1^\beta\ofx e_2^\gamma\ofx \epsilon_{ijk}\left[\partial_jn^\beta\ofx\right]\left[\partial_kn^\gamma\ofx\right].\end{equation}
We see then that the curl of ${\bm{\Omega}}\ofx$ does not always vanish, but it
appears that it depends on our arbitrary vectors ${\bf e}_1\ofx$ and ${\bf e}_2\ofx$.  Note however
that if we interchange the indices $\beta$ and $\gamma$ in (\ref{ecurlofomegathree}) that
it is equivalent to interchanging $j$ and $k$, which would introduce a minus sign.  Thus we
have
\begin{eqnarray}\label{emh}
\left[\nabb\cross{\bm{\Omega}}\ofx\right]_i &=&
\frac{1}{2}\left[e_1^\beta\ofx e_2^\gamma\ofx - e_1^\gamma\ofx e_2^\beta\ofx\right]
\epsilon_{ijk}\left[\partial_jn^\beta\ofx\right]\left[\partial_kn^\gamma\ofx\right]\nonumber\\
&=& \frac{1}{2}\epsilon_{\alpha\beta\gamma}n^\alpha\ofx \epsilon_{ijk}\partial_j n^\beta\ofx
\partial_kn^\gamma\ofx,\end{eqnarray}
where we have used the orthonormality of $\{{\bf e}_1\ofx,{\bf e}_2\ofx,{\bf n}\ofx\}$.  This is
the celebrated Mermin-Ho relation \cite{MH}.  Note that because of the antisymmetry
of $\epsilon_{ijk}$, it is unnecessary to keep the brackets around the gradients of 
${\bf n}\ofx$.

This relation between ${\bf n}\ofx$ and vectors perpendicular to ${\bf n}\ofx$ is rather remarkable.  Note that our choice of ${\bf
e}_1\ofx$ and ${\bf e}_2\ofx$ is arbitrary so that
${\bm{\Omega}}\ofx$ is not a constant of the geometry.  However,
(\ref{emh}) shows that ${\bm{\nabla}}\times{\bm{\Omega}}\ofx$ only
depends on ${\bf n}\ofx$, and not our choice of basis vectors!
This is why the Mermin-Ho relation has been introduced in this
section on global properties \cite{LWD}: if we have a closed curve
$\Gamma$ of length $L$, then
\begin{eqnarray}\label{eclosed}
\oint_\Gamma \big[{\bm{\nabla}}\theta\ofx -
{\bm{\Omega}}\ofx\big]\cdot d{\bf R}&=&
\left[\theta(L)-\theta(0)\right] - \dint_M
\left[\nabb\times{\bm{\Omega}}\ofx\right]\cdot d{\bf S}\nonumber\\
&=& \left[\theta(L)-\theta(0)\right] - \dint_M
\frac{1}
{2}\epsilon_{\mu\nu\rho}\epsilon_{\alpha\beta\gamma}n^\alpha\ofx\partial_\nu
n^\beta\ofx\partial_\rho n^\gamma\ofx dS_\mu\end{eqnarray}
where we have
used Stokes theorem to change an integral around a curve $\Gamma$
into an integral over a capping surface $M$.  In (\ref{eclosed}) $d{\bf
S}$ is an element of area.  Thus if we consider changes in a
vector around a closed curve, our choice of ${\bf e}_i\ofx$ is
irrelevant.  Though it is necessary that ${\bf N}(L)={\bf N}(0)$,
$\theta(L)-\theta(0)$ can change by an integral multiple of
$2\pi$.  The rest of the change in the direction of ${\bf N}\ofx$
comes from the geometry of ${\bf n}\ofx$.  We will see how this
extra term plays an important role in the next section, in the
physics of surfaces and the physics of defects in three
dimensions.

\subsection{Link, Twist and Writhe: {\sl Dynamics of twist-storing polymers}}

One of the more interesting stiff polymers is DNA.  Though it is well
known that it has great biological significance \cite{watson}  , it is of interest
in materials for {\sl at least} two other reasons.  First, the persistence length
of DNA is quite long, roughly $50$ nm.  More importantly, because there are
a plethora of enzymes available to cleave DNA, a sample of monodisperse
polymers can be prepared much more readily than in any synthetic system.
We have already discussed conformations of chiral polymers like DNA in \S{\bf II}.  However
DNA has one more interesting element: it can form into a closed loop.  This
is interesting because DNA is actually a double stranded helix.  Therefore
the number of times that one strand wraps around the other is fixed when the loop
is closed (when the strand is open the helix can unravel).  Thus there is a conserved
quantity and this leads to a constraint on the possible dynamics of
the double strand.  More generally, any polymers that cannot unwind along their axis
are known as twist-storing polymers.

When we have two closed curves $\Gamma$ and $\Gamma'$,
we can assign a linking number \Lk\ to them
which counts the number of times one loop passes through the other.  There is a simple
way to calculate this given the two curves $\vecR(s)$ and $\vecR'(s)$ using an analogy
with Amp\`ere's law.  If we think of the first curve $\Gamma$ as being a wire carrying a
current $I$, then we know that the line integral of the generated magnetic field ${\bf B}$ around
the closed curve $\Gamma'$ is $4\pi n I$, where $n$ is the number of times
that the current passes through the closed loop\footnote{For simplicity we work
in cgs units with $c=1$}.  Setting $I=1$, we have
\begin{equation}\label{eampere}
\Lk =  n = \frac{1}{4\pi}\oint_{\Gamma'} {\vecB}\ofxp\cdot d{\bf x}'.\end{equation}
We may calculate the resulting field ${\bf B}\ofxp$ from the wire by use of the
Biot-Savart Law:
\begin{equation}\label{ebiotsavart}
\vecB\ofxp = \oint_\Gamma \frac{Id{\bf\ell}\cross {\hat {\bf r}}}{r^2} =
\oint_\Gamma \frac{d{\bf x}\cross\left[{\bf x}'-{\bf x}\right]}{\vert {\bf x}'-{\bf x}\vert^3}.\end{equation}
Putting these together we find that (with the curves having length $L$ and $L'$, respectively):
\begin{eqnarray}\label{egaussinv}
\Lk &=& {1\over 4\pi}\oint_\Gamma \oint_{\Gamma'}  {\left[{\bf x}-{\bf x}'\right]\cdot\left(d{\bf x}\cross
d{\bf x}'\right)\over \vert {\bf x}-{\bf x}'\vert^3} \nonumber\\
&=& {1\over 4\pi} \oint_0^L ds \oint_0^{L'} ds' \left[{d\vecR(s)\over ds}\cross{d\vecR'(s')\over ds}\right]\cdot{\left[\vecR(s)-\vecR'(s')\right]\over \vert \vecR(s) - \vecR'(s')\vert^3}\nonumber\\
&\equiv& G(\Gamma,\Gamma')\end{eqnarray}
the last equality defines $G(\Gamma,\Gamma')$, the {\sl Gauss Invariant}.
Thus two closed curves that cannot separate and rejoin must keep this invariant
constant.

While this may be elegant, it is not especially useful when studying molecules like DNA.  At
long lengthscales, DNA appears as a single filament.  It would be useful to recast the
linking number in terms of the single polymer picture of DNA.  To do this, we
consider two curves that are close together \cite{fuller}.  The first curve is
$\vecR(s)$, while the second curve is
\begin{equation}\label{esecondcurve}\vecR'(s') = \vecR(s') + \epsilon {\bf u}(s')\end{equation}
where, as is the tradition, $\epsilon$ is a small number and ${\bf u}(s')$ is a unit
vector that is perpendicular to the tangent vector ${d\vecR(s')/ds}=\vecT(s')$.  Inserting
these expressions into (\ref{egaussinv}), we have
\begin{equation}\label{elktwwr}
\Lk = {1\over 4\pi} \oint_0^L \!\!ds \oint_0^{L'}\!\! ds' \,\vecT(s)\cross\left[\vecT(s') + \epsilon{d{\bf u}(s')\over ds}\right]\!\cdot\!{\left[\vecR(s)-\vecR(s') - \epsilon{\bf u}(s')\right]\over \vert \vecR(s) - \vecR(s') - \epsilon{\bf u}(s')\vert^3}.\end{equation}
We now want to take $\epsilon$ to zero.  While the numerator in (\ref{elktwwr}) does not make
this a problem, note that the denominator diverges whenever $s=s'$.  Thus as long
as $\vert s-s'\vert \ge \delta$ we can take $\epsilon\rightarrow 0$ \cite{kleinertbook}.   We thus have
\begin{eqnarray}\label{elktw}
\Lk &=& {1\over 4\pi} \oint_0^{L'} ds' \left\{\int_0^{s'-\delta}ds + \int_{s'+\delta}^{L} ds\right\}\,
{\vecT(s)\cross\vecT(s')\cdot\left[\vecR(s)-\vecR(s')\right]\over \vert\vecR(s)-\vecR(s')\vert^3}\nonumber\\
&&\;\; + {1\over 4\pi}\oint_0^{L'}\!\!\! ds' \int_{s'-\delta}^{s'+\delta} \!\!ds\, \vecT(s)\cross\left[\vecT(s') + \epsilon{d{\bf u}(s')\over ds}\right]\cdot{\left[\vecR(s)-\vecR(s') - \epsilon{\bf u}(s')\right]\over \vert \vecR(s) - \vecR(s') - \epsilon{\bf u}(s')\vert^3}\end{eqnarray}
The first integral bears a resemblance to the Gauss invariant, while the
second integral depends on the vector ${\bf u}(s)$.  Since $\delta$ is small, we
can expand $\vecR(s)$ around $s'$ to find:
\begin{equation}\label{eexpand}\vecR(s) = \vecR(s') + (s-s')\vecT(s') + \ldots\end{equation}
and
\begin{equation}\label{etexpand}\vecT(s) = \vecT(s') + (s-s'){d\vecT(s')\over ds} +\ldots\end{equation}
Inserting this into the second integrand and using the fact that
$\vecT(s)\cdot{\bf u}(s)=0$, we find
\begin{eqnarray}\label{elktww}
\Lk &=& {1\over 4\pi} \oint_0^L ds\oint_0^{L'} ds'\,\vecT(s)\cross\vecT(s')\cdot{
\vecR(s)-\vecR(s')\over\vert\vecR(s)-\vecR(s')\vert^3}\nonumber\\
&&\;\; - {1\over 4\pi} \oint_0^{L'} \!\!\!ds' \int_{s'-\delta}^{s'+\delta} \!\!ds\, {\left[\vecT(s')+(s-s'){d\vecT(s')\over ds} \right]\cross\left[
\vecT(s') + \epsilon{d{\bf u}(s')\over ds}\right]\cdot\epsilon{\bf u}(s')
\over \left[(s-s')^2+\epsilon^2\right]^{3/2}}
\end{eqnarray}
where we have taken $\epsilon\rightarrow 0$ in the first integral and have
used $\vecT(s')\cross\vecT(s')=0$ in the second.  Note that the $s$ dependence
of the second integral is manifest and we may do the integration over $s$, let $\epsilon$
go to zero {\sl first} and then let $\delta$ go to zero.   We have separated
the link into two integrals.  The first is called the {\sl writhe}, \Wr , and it only
depends on the curve $\vecR(s)$.  The second is called the {\sl twist}, \Tw , and
it depends on ${\bf u}(s)$.  We have
\begin{eqnarray}\label{etwwra}
\Wr & = &{1\over 4\pi}  \oint_0^L ds\oint_0^L ds'  \,\vecT(s)\cross\vecT(s')\cdot{\vecR(s)-\vecR(s')\over
\vert\vecR(s)-\vecR(s')\vert^3}\\ \label{etwwrb}
\Tw & =& {1\over 2\pi}\oint_0^L ds \,\vecT(s)\cdot\left[{\bf u}(s)\cross{d{\bf u}(s)\over ds}\right]\end{eqnarray}
and thus we arrive at Fuller's celebrated result \cite{fuller}:
\begin{equation}\label{efullers}\Lk = \Tw + \Wr.
\end{equation}
Though the expression for writhe (\ref{etwwra}) bares a strong resemblance 
to that for the Gauss invariant (\ref{egaussinv}), it is not the same.  In the Gauss invariant
we were considering two different curves that did not touch and there was no need
to expunge a singularity from the integration.  The expression for writhe, on the other hand, is
for the same curve and is only defined in the limit described above.
Note further that the writhe is nonlocal, while the twist is local.  This is the price we must pay: the
link is topological, the writhe depends only on the backbone $\vecR(s)$ but is nonlocal, and
the twist is local but we must know about ${\bf u}(s)$, {\sl i.e.} the other curve.

\vskip10pt
\begin{figure}\epsfxsize=2.5truein
\epsfbox{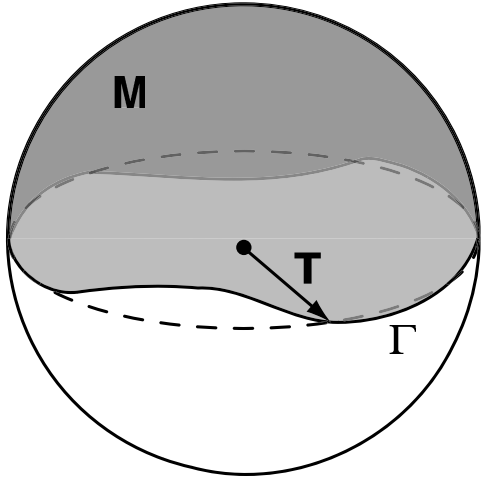}
\vskip10pt
\caption{The tangent spherical map.  The area of the patch $M$ on the
sphere is, through Stokes theorem, the line integral around $\Gamma$ of 
${\bf e}_1\cdot\nabla{\bf e}_2$.}
\end{figure}
\vskip10pt

Note that the expression for the twist (\ref{etwwrb}) suggests a use of the results of \S{\bf III}B.  Since
${\bf u}(s)$ is perpendicular to $\vecT(s)$, we may expand it in a set of basis vectors ${\bf e}_1(s)$
and ${\bf e}_2(s)$ both perpendicular to $\vecT(s)$, so that ${\bf u}(s)=\cos\theta(s){\bf e}_1(s) + \sin\theta(s){\bf e}_2(s)$.  Then
\begin{equation}\label{etwistt}\vecT(s)\cdot\left[{\bf u}(s)\cross{d{\bf u}(s)\over ds}\right] = \partial_s\theta(s) -
{\bf e}_1(s)\cdot\partial_s{\bf e}_2(s)\end{equation}
where we have used the fact that ${\bf e}_1\cross{\bf e}_2=\vecT$ (plus cyclic permutations).
But this is the form we considered in the discussion of the change in a vector.  Following
the argument at the end of the previous section, we find that
\begin{eqnarray}\label{etwsis}
\Tw &=& {1\over 2\pi}\oint_0^L ds\,  \left[\partial_s\theta(s) - {\bf e}_1(s)\cdot\partial_s{\bf e}_2(s)\right] \nonumber\\
&=&
m -  \dint_M
{1\over
4\pi}\epsilon_{\mu\nu\rho}\epsilon_{\alpha\beta\gamma}T^\alpha\ofx\partial_\nu
T^\beta\ofx\partial_\rho T^\gamma\ofx dS_\mu\nonumber\\
&=&\Lk - \Wrp,\end{eqnarray}
where $m$ is an integer (since $\theta(L)-\theta(0)=2\pi m$).  We would be tempted
to identify the second integral $\Wrp$ as the writhe, since we have an equation that
reads justs as (\ref{efullers}).  This is not always correct and depends on our
choice of basis vectors.  However, it does establish a weaker result of Fuller's \cite{Fullerii} that
the writhe is the last term in (\ref{etwsis}) {\sl mod} 1.  As we will see in the next section,
the integral has a geometric meaning: it is the area swept out by the tangent curve on
the tangent spherical map, as shown in Figure 3.

The difficulty with writhe is that it is non-local and therefore cannot be easily
included in a local set of dynamical laws -- the conformation of the entire curve
must be known to calculate writhe.  Moreover, writhe is only defined for closed curves and
so the identity (\ref{efullers}) would not appear to apply to open strands.  
While the total amount of writhe including the integral part is important
for calculating ground states of a particular twisted ribbon
\cite{markosiggia,Rudnick, Rudnickii,Julicher} when considering changes in writhe, the constant integer part is less important.  
If the timescale for the diffusion of twist along the polymer is long enough, then
even an open strand should feel the constraint of conserved link.  

To this end, a useful
result for the change in writhe of a curve
as a function of time $t$ \cite{knot}:
\begin{equation}\label{euseful}{\partial_t \Wr(t)} = {1\over 2\pi}
\oint ds\, {\bf T}(s,t)
\inner\left[\partial_t{\bf T}(s,t)\cross\partial_s{\bf T}(s,
t)\right],\end{equation}
can be used.  Note that this result follows from the preceding discussion: if 
we choose a coordinate system, then
(\ref{euseful}) (multiplied by $dt$) is the differential of area swept out
by the tangents of
two closed curves with tangent vectors
${\bf T}(s,0)$ and ${\bf T}(s,dt)$.  Since the twist is really
the local torsional strain of the polymer, we denote the twist density as $\omega(s,t)$ and
then we have
\begin{equation}\label{ederiv}\partial_t \Lk = \oint ds\,\left\{\partial_t\omega(s,t)
+ \partial_t{\bf T}(s,t)\inner\left[\partial_s{\bf T}(s,t)\cross{\bf T}(s,t)\right]\right\}+
\partial_t n\end{equation}
where $n$ is the integer difference between $\Wr$ and $\Wrp$.  Since continuous
evolution
cannot lead to discontinuous changes in the integer $n$ and since link is
conserved we have
\begin{equation}\label{edrri}0=\oint ds\,\left\{\partial_t\omega
+ \partial_t{\bf T}\inner\left[\partial_s{\bf T}\cross{\bf T}\right]\right\}.\end{equation}
This conservation law need not be satisfied locally: the curve can twist in one
place and writhe at some distant location to satisfy (\ref{edrri}) .  In
addition, the integer can change if the curve develops cusps and
evolves in a non-smooth way.   Because physics is local, however, we
might expect that linking number is locally conserved and changes
via a ``link current'' $j$.  This would lead to a 
local conservation law
\begin{equation}\label{elinck}\partial_s j = \partial_t\omega + \partial_t{\bf
T}\inner\left[\partial_s{\bf T}
\cross{\bf T}\right],\end{equation}
which satisfies (\ref{edrri}) .  This conservation law enforces total
link conservation since $\oint ds\,\partial_s j\equiv 0$, yet allows
for {\sl local} deviations in the twist and writhe \cite{LWD} .

This geometrically inspired conservation law has been verified in numerical
experiments \cite{twirlwhirl,twirlwhirlii} which show that there are two modes of relaxation
in the dissipative dynamics: {\sl twirling}, in which the link relaxes through torsional
modes along the polymer (like a speedometer cable), and whirling, in which the
link relaxes through crank-shaft like motions of the entire chain.

\section{Local Theory of Surfaces}

\subsection{The Area Element: {\sl Minimal Surfaces}}

\vskip10pt
\begin{figure}\epsfxsize=3.5truein\epsfbox{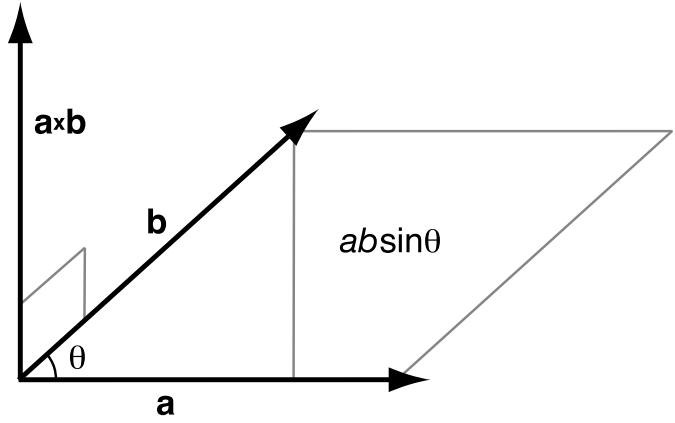}
\vskip10pt
\caption{The magnitude of the cross product of two vectors is the
area of the parallelogram swept out by them.}
\end{figure}
\vskip10pt

While it is easy to generalize the equation of a curve $\vecR(s)$ to an equation
for a surface $\vecX(s_1,s_2)$, it turns out that the theory of surfaces is much
richer than that for curves.  In the first place, there is no way to talk about arclength
when defining $s_1$ and $s_2$, so we will take them to be arbitrary coordinates
of the surface.  We can, however, consider the area of a small patch of our
surface.  To do this, we need to construct vectors tangent to the surface.  As shown in
Figure 4, if we have
two vectors ${\bf a}\ofss$ and ${\bf b}\ofss$ in the surface, then they span a surface area
$\Delta S=\vert {\bf a}\ofss\vert\;\vert {\bf b}\ofss\vert \sin\theta$, where $\theta$ is the angle between the two
vectors.  We recognize this expression for $\Delta S$ as the magnitude of the
cross product ${\bf a}\ofss\cross{\bf b}\ofss$.  Since both of these vectors are tangent to the surface,
their cross product is parallel to the unit layer normal ${\bf n}\ofss$.  Thus, we have
\begin{equation}\label{earea}\Delta S = \left\vert {\bf n}\ofss\cdot\left[{\bf a}\ofss\cross{\bf b}\ofss\right]\right\vert\end{equation}
But, with this construction we can also construct the unit layer normal out of
${\bf a}\ofss$ and ${\bf b}\ofss$:
\begin{equation}\label{elaynorm}{\bf n}\ofss = \pm{{\bf a}\ofss\cross{\bf b}\ofss\over\vert\vert{\bf a}\ofss\cross{\bf b}\ofss\vert\vert}.\end{equation}
Putting this expression together with (\ref{earea}), we have
\begin{eqnarray}\label{ethearea}
\Delta S &=& {\left[{\bf a}\ofss\cross{\bf b}\ofss\right]\cdot\left[{\bf a}\ofss\cross{\bf b}\ofss\right]\over \vert\vert{\bf a}\ofss\cross{\bf b}\ofss\vert\vert}\nonumber\\
&= &\sqrt{\big[{\bf a}\ofss\cross{\bf b}\ofss\big]^2}\nonumber\\
&=& \sqrt{{\bf a}^2\ofss{\bf b}^2\ofss - \big({\bf a}\ofss\cdot{\bf b}\ofss\big)^2}\end{eqnarray}
We now need to construct two, nonparallel vectors ${\bf a}\ofss$ and ${\bf b}\ofss$.  The surface
provides us with two: ${\bf a}\ofss = \partial_{s_1}\vecX\ofss ds_1$ and ${\bf b}\ofss = \partial_{s_2}\vecX\ofss ds_2$.  This tells us that the differential area
element is;
\begin{equation}\label{earelem}dS = \sqrt{\big(\partial_{s_1}\vecX\ofss\big)^2\big(\partial_{s_2}\vecX\ofss\big)^2 -
\big(\partial_{s_1}\vecX\ofss\cdot\partial_{s_2}\vecX\ofss\big)^2}\;ds_1 ds_2.\end{equation}
Often we also need the vector surface element $d{\bf S}= {\bf n}\ofss dS$, for instance when
calculating electric and magnetic flux.
The area of the whole surface $M$ is simply
\begin{equation}\label{eallarea}A = \int_M \sqrt{\big(\partial_{s_1}\vecX\big)^2\big(\partial_{s_2}\vecX\big)^2 -
\big(\partial_{s_1}\vecX\cdot\partial_{s_2}\vecX\big)^2}\,ds_1ds_2\end{equation}
What happens if we choose to parameterize the surface in terms of a new
set of coordinates?  If we have two new coordinates $\sigma_1$ and $\sigma_2$
defined by $\sigma_i=\sigma_i\ofss$, then by the chain rule
\begin{equation}\label{echange}\partial_{s_i}\vecX={\partial\vecX\over\partial s_i}={\partial\vecX\over\partial\sigma_j}{\partial\sigma_j\over\partial s_i}={\partial\sigma_j\over\partial s_i}\partial_{\sigma_j}\vecX,\end{equation}
where, as usual, there is an implicit sum over $j$.  Because of this sum, it
appears that it would be somewhat tedious to reexpress (\ref{eallarea}) in terms of
the new coordinates.  However, we note that if we define a matrix ${\bf g}$:
\begin{equation}\label{emetric}{\bf g}\ofss = \left[\begin{matrix}\partial_{s_1}\vecX\cdot\partial_{s_1}\vecX&&\partial_{s_1}\vecX\cdot\partial_{s_2}\vecX\\ \\
\partial_{s_2}\vecX\cdot\partial_{s_1}\vecX&&\partial_{s_2}\vecX\cdot\partial_{s_2}\vecX\\ \end{matrix}\right],\end{equation}
then the expression in the radical of (\ref{eallarea}) is simply
the determinant of this matrix.
In component form, this matrix is
$g_{ij}\ofss=\partial_{s_i}\vecX\ofss\cdot\partial_{s_j}\vecX\ofss$.
This matrix is known as the {\sl metric tensor} which some people find useful \cite{MTW}.   Moreover, we note that the transformation (\ref{echange}) amounts to
matrix multiplication of ${\bf g}\ofss$:
\begin{equation}\label{emetricc}g_{ij}\ofss = {\partial\sigma_k\over\partial s_i}{\partial\sigma_m\over\partial s_j}
\widetilde g_{km}(\sigma_1,\sigma_2),\end{equation}
where $\widetilde g_{km}(\sigma_1,\sigma_2)$ is the corresponding matrix in the
new coordinate system.   Defining a new matrix ${\bf O}$ by
$O_{ij} = {\partial\sigma_j\over\partial s_i}$, then ${\bf g} = {\bf O}^T\widetilde{\bf g}{\bf O}$.
Thus when we change coordinates, we have
\begin{equation}\label{echangeii}A = \int_M \sqrt{\det{\bf g}}\,ds_1ds_2 = \int_M \sqrt{\det{\bf O}^T\widetilde{\bf g}{\bf O}}\,
ds_1 ds_2 = \int_M\sqrt{\det\widetilde{\bf g}} \,\,\vert\!\det{\bf O}\vert\, ds_1 ds_2.\end{equation}
But the determinant of ${\bf O}$ is just the Jacobian determinant of the transformation from
$s_i$ to $\sigma_i$, and so $\vert\!\det{\bf O}\vert\,ds_1ds_2 = d\sigma_1 d\sigma_2$ or
\begin{equation}\label{einvariant}
A = \int_M \sqrt{\det{\bf g}}\, ds_1 ds_2 = \int_M\sqrt{\det{\widetilde{\bf g}}}\, d\sigma_1 d\sigma_2\end{equation}
and so the area is invariant under coordinate transformations.  This invariance (known
as {\sl diffeomorphism covariance} by the {\sl cognoscenti}) is useful as it allows us
to choose the most convenient set of coordinates for a given problem.  One often used
choice is the so-called {\sl Monge gauge} or {\sl height representation},
where $s_1$ and $s_2$ are chosen to
be the $x$ and $y$ components of the surface vector $\vecX\ofss$ and the $z$-component
is a function of $x$ and $y$:
\begin{eqnarray}\label{emonge}
x=X_1\ofss&=&s_1\nonumber\\
y=X_2\ofss&=&s_2\nonumber\\
z=X_3\ofss&=&h(s_1,s_2)=h(x,y)\end{eqnarray}
It is straightforward to check that the expression for the area (\ref{eallarea}) becomes the more
familiar expression:
\begin{equation}\label{emongearea}A = \int_M \sqrt{1 + \left({\partial h\over\partial x}\right)^2 + \left({\partial h\over\partial y}\right)^2}\,dxdy\end{equation}
The height representation requires that for every $x$ and $y$, there is only one height or,
in other words, the surface can have no overlaps.  Sometimes it is necessary to break
the surface up into different regions or {\sl patches} in order to make a good height
representation.

The height representation proves useful when considering the energetics of fluid membranes.  By
{\sl fluid} we mean that the membrane has no internal structure and the molecules in
it can flow freely.  Often we are interested in the shape and dynamics of these membranes
when they are under a uniform applied tension.  Soap films are a common example, though
in many cases the interface between two fluids or two distinct phases can also
be thought of as a membrane under tension.  The free energy of this system is simply
$F= \varsigma A$, where $\varsigma$ is the surface tension and $A$ is the area
of the interface.   Minimizing the free
energy thus amounts to minimizing the area of the surface.  The expression for
the area in (\ref{emongearea}) looks like an action from classical mechanics.  If we
define the Lagrange density ${\cal L}$:
\begin{equation}\label{emonlag}{\cal L} = \sqrt{1 + \left({\partial h\over\partial x}\right)^2 + \left({\partial h\over\partial y}\right)^2},\end{equation}
then the Euler-Lagrange equation is \cite{goldmech}
\begin{equation}\label{eeullag}{\partial\over \partial x} {\partial{\cal L}\over\partial(\partial_x h)}
+{\partial\over\partial y}{\partial{\cal L}\over\partial(\partial_y h)} = {\partial{\cal L}\over \partial h},\end{equation}
or, by inserting (\ref{emonlag})
\begin{equation}\label{eminim}{\partial\over\partial x}\left({\partial_x h\over \sqrt{1+(\nabb h)^2}}\right)
+ {\partial\over \partial y}\left({\partial_y h\over\sqrt{1 + (\nabb h)^2}}\right)=0.\end{equation}
This equation has an important interpretation in terms of the surface normal.  If we construct
the two tangent vectors to the surface as we did above, we have
\begin{eqnarray}\label{twotangents}
{\bf u}_1 &= \partial_{s_1}\vecX\ofss=\left[1,0,\partial_xh\right]\nonumber\\
{\bf u}_2 &=\partial_{s_2}\vecX\ofss=\left[0,1,\partial_yh\right]\end{eqnarray}
and so the surface normal is parallel to ${\bf u}_1\cross{\bf u}_2=\left[-\partial_x h,-\partial_y h, 1\right]$.  Normalizing this vector gives us ${\bf n}\ofss = -\left[\partial_x h,\partial_y h, -1\right]/\sqrt{1+(\nabb h)^2}$ and so (\ref{eminim}) becomes $\nabb\!\cdot\! {\bf n}=0$.  In other
words, a surface which extremizes its area has a divergence-free unit normal.  We will
address this further in the following section. Such
a surface is called a {\sl minimal surface}.  Typically, (\ref{eminim}) is written equivalently as \cite{Nitsche}
the
minimal surface equation:
\begin{equation}\label{eminimal}\left[1+(\partial_yh)^2\right]\partial^2_xh - 2\partial_xh\,\partial_yh\,\partial_x\partial_yh + \left[1+(\partial_xh)^2\right]\partial^2_yh=0.\end{equation}

\vskip10pt
\begin{figure}\epsfxsize=4.5truein\epsfbox{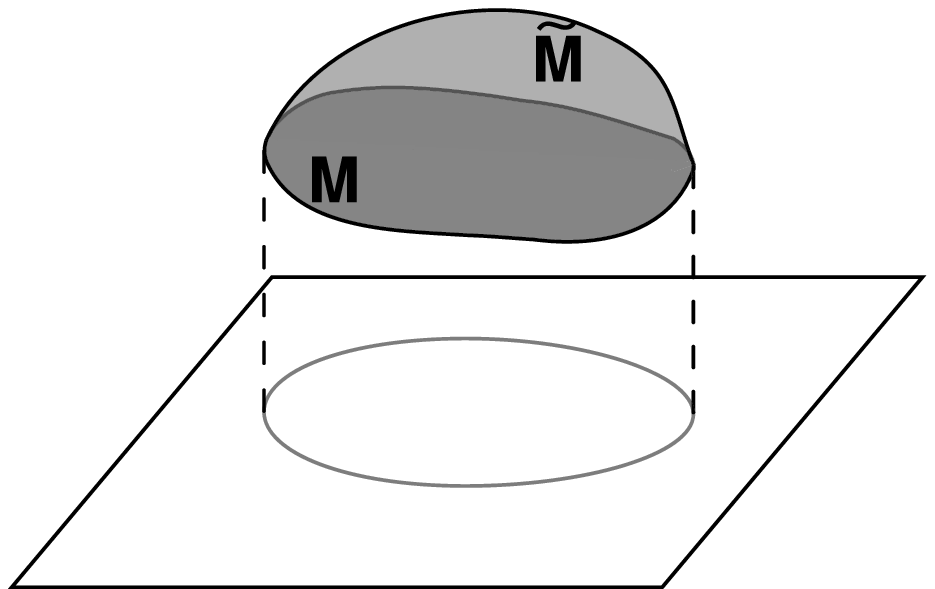}
\vskip10pt
\caption{The original minimal surface ${\bf M}$ and the comparison
surface $\widetilde{\bf M}$.}
\end{figure}
\vskip10pt

Though we were seeking a surface with a minimum area, the Euler-Lagrange equations
only guarantee an extremum -- we might have a saddle-point or, worse, a maximum!  Rigorous
proofs that the area is minimized in general are difficult.  However, we can prove a
restricted result easily.  If the projection of the boundary $\partial M$ onto the $xy$-plane is convex, then it is clear
that any overhangs on the surface add extra area and that the surface need
not ``stick out'' past the boundary of $M$ projected onto the $xy$-plane.  Thus we can use the height representation and solve (\ref{eminimal}) subject to a boundary condition
$h(x,y) = f(x,y)$ on the boundary in the $xy$-plane.
As depicted in Figure 5, we consider an arbitrary surface $\widetilde M$ with the {\sl same boundary} as the surface of which we are
interested.  Since the same range of $x$ and $y$ map out both $M$ and $\widetilde M$, we
can define ${\bf n}(x,y)$ on $\widetilde M$ and in the volume between the surfaces. Define
\begin{equation}\label{eT}T = \int_{\widetilde M} {\bf n}\ofx\!\cdot\! d\widetilde{\bf S}\end{equation}
where ${\bf n}\ofx$ is the unit normal of the original surface $M$, and $d\widetilde{\bf S}$ is the vector
element of surface area, pointing along the normal to the surface $\widetilde{\bf n}\ofx$.  
If we consider the
same integral for the original surface $M$, then since ${\bf n}\ofx$ is parallel to $d{\bf S}$, we
have ${\bf n}\ofx\cdot d{\bf S}=dS$ the element of surface area! Integrating this over the whole surface just gives us the area $A$.  Thus
\begin{equation}\label{eA}A = \int_M dS = \int_M {\bf n}\ofx \!\cdot\! d{\bf S}.\end{equation}
Moreover, we have the inequality $\tilde{\bf n}\ofx\!\cdot\!{\bf n}\ofx\le 1$, since the
dot-product of two unit vectors is the cosine of some angle and is therefore less than or
equal to 1.  We arrive at:
\begin{equation}\label{eAT}T=\int_{\tilde M}{\bf n}\ofx\!\cdot\! d\tilde{\bf S} = \int_{\tilde M} {\bf n}\ofx\!\cdot\!\widetilde{\bf n}\ofx
d\tilde S \le \int_{\widetilde M} d\widetilde S = \widetilde A,\end{equation}
where $\tilde A$ is the area of the comparison surface.  We are almost done.  Note
that if we construct a closed surface $\widehat M$ by gluing together $M$ and $\widetilde M$ along their
common boundary, then the integral of ${\bf n}\ofx$ over the whole surface
is:
\begin{equation}\label{ewhole}\int_{\widehat M} {\bf n}\ofx\!\cdot\! d\widehat{\bf S} = \int_M{\bf n}\ofx\!\cdot\! d{\bf S} -
\int_{\widetilde M} {\bf n}\ofx\!\cdot\! d\widetilde{\bf S} = A - T\end{equation}
where the relative minus sign arises because we want the normal to the
closed surface to always point outward.  But by Gauss's law, the first integral in (\ref{ewhole}) can
be converted into an integral over the volume $\widehat V$ enclosed by $\widehat M$, so
\begin{equation}\label{egaussslaw}
A-T = \int_{\widehat M}{\bf n}\ofx\!\cdot\! d\widehat{\bf S} = \int_{\widehat V} \nabb\!\cdot\!{\bf n}\ofx \,d\widehat V =0,\end{equation}
since $M$ extremizes the area so $\nabb\!\cdot\!{\bf n}\ofx=0$.  Using the inequality (\ref{eAT}) we
have $A=T\le \widetilde A$ and thus the comparison surface has an area at least as large
as the original surface $M$.  Since $\widetilde M$ was arbitrary we have shown that
our surface $M$ minimizes the area\footnote{In this context, the vector field ${\bf n}\ofx$ is
called a {\sl calibration}.  The interested (or disinterested) reader will find this and other
results in \cite{Morgan}.}.
\vskip10pt
\begin{figure}\epsfxsize=4.5truein\epsfbox{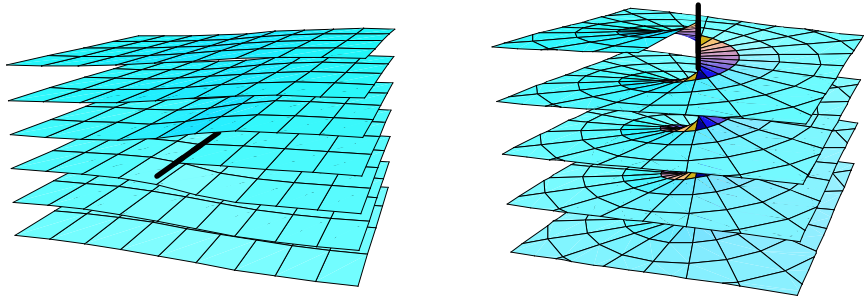}
\vskip10pt
\caption{Edge (left) and screw (right) dislocations in a smectic-$A$ liquid crystal.}
\end{figure}
\vskip10pt

Examples of minimal surfaces abound.  The smectic-$A$ liquid crystal phase is
a one-dimensional crystal, composed of periodic layers of spacing $d$. Each layer
is fluid-like, with no internal ordering.  Thus, in some sense, each layer
behaves as a minimal surface and the layer spacing is set by a compression
modulus \cite{KLScherk}.  The smectic phase can have topological defects of two
types.  As shown in Figure 6, an {\sl edge dislocation} adds or ends a layer in the bulk, while a {\sl screw dislocation} connects adjacent layers together.  A screw defect can be
represented by the height function:
\begin{equation}\label{escrew}h(x,y) = {nd\over 2\pi}\tan^{-1}\left(y\over x\right)\end{equation}
where $n$ is an integer.  By choosing $n$ to be an integer, we are guaranteed that
when going around the origin once we move up exactly $n$ layers.  It is straightforward
to check that the screw dislocation is a minimal surface by direct calculation of (\ref{eminimal}).

Another nice example also arises in smectic liquid crystals.  Scherk's first surface (shown in Figure 7) \cite{Scherk} is a minimal surface which smoothly connects two smectic
regions which are rotated with respect to each other.  It is composed of an infinite set
of parallel, screw-like dislocations \cite{elt,aml} which must be slightly squashed to
make the surface satisfy (\ref{eminimal}).  For those who are interested, the height
function is defined for arbitrary rotation angle $\alpha$:
\begin{equation}\label{scherk}h[x,y;\alpha]\equiv-\sec({1\over 2}\alpha)
\tan^{-1}\left\{{
\tanh\left[{1\over 2}x\sin(\alpha)\right]\over
\tan\left[y\sin({1\over 2}\alpha)\right]}\right\}.\end{equation}
It is interesting to note that as $\alpha\rightarrow 0$ the height function becomes
that of a simple screw dislocation (\ref{escrew}) with $nd=2\pi$.

\vskip10pt
\begin{figure}\epsfxsize=4.5truein\epsfbox{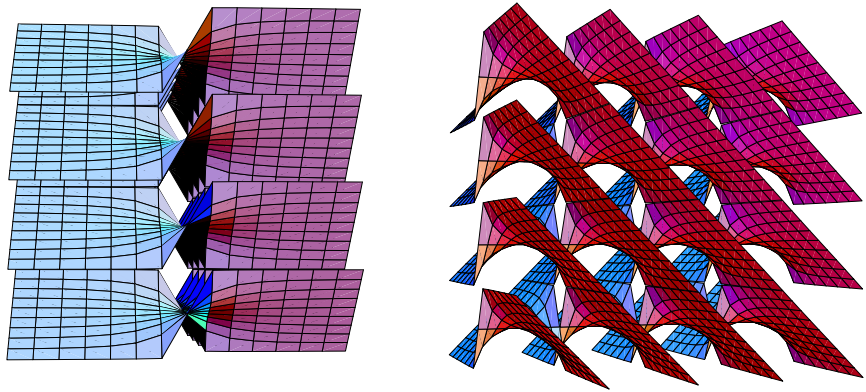}
\vskip10pt
\caption{Scherk's first surface.  This minimal surface connects together two
layered structures with
different orientations.}
\end{figure}
\vskip10pt

\subsection{Mean and Gaussian Curvature: {\sl Energetics of Membranes}}

When we discussed curves in three dimensions there were not many options
for the definition of curvature: there was only one coordinate on the curve, and if
we wanted something that was independent of our parameterization, we need only
take derivatives with respect to arclength.  As we have seen in the last section, however,  surfaces
can be reparameterized and have many more degrees of freedom.  As a result, there
is more than one way to define curvature, in fact there are two.

We have already hinted at one definition in the discussion of minimal surfaces.   For curves, we
know that a straight line minimizes the distance between two points in space, or in
other words, $\kappa(s)=0$ is the equation of a length-minimizing curve.  Generalizing
this to surfaces, we might define one curvature which is $0$ for surfaces that minimize
the area spanned by a given boundary.  Indeed, this is called the {\sl mean curvature}, $H$
and $H=-\frac{1}{2}\nabla\!\cdot\!{\bf n}\ofx$ where ${\bf n}\ofx$ is the surface normal.  The other
type of curvature is the {\sl Gaussian curvature}, $K$, which we will define in the following.

In order to understand this better and to introduce $K$, we need to step back and
consider our surface $\vecX\ofss$.  Since the curvature of a line is $\kappa(s) = - \vecT(s)\cdot\vecN'(s)$, we would like to
take derivatives of the surface normal along tangent directions of the surface.  As in \S{\bf III}B, we now reintroduce the
basis vectors perpendicular to $\vecn\ofss$, ${\bf e}_1\ofss$ and ${\bf e}_2\ofss$, of unit length and mutually
orthogonal.  We can take directional derivatives\footnote{It may seem odd to use the gradient in these definitions since
$\vecn\ofss$ is a function of $s_1$ and $s_2$.  The derivatives of $\vecn\ofss$ along the surface tangent vectors
$\partial_{s_j}\vecX\ofss$ are just $\partial_{s_j}\vecn\ofss$.  By writing ${\bf e}_i\ofss$ as the linear
combination ${\bf e}_i=A_{ij}\partial_{s_j}\vecX$, the chain rule implies that
$\left({\bf e}_i\cdot\nabb\right) = A_{ij}\partial_{s_j}$.} of $\vecn\ofss$ along ${\bf e}_i\ofss$,
$\left({\bf e}_i\cdot\nabb\right)$.  Now instead of one number, we have four, which form a matrix:
\begin{equation}\label{esecondnorm}{\bf L} =-\left[\begin{matrix}{\bf e}_1\cdot\left[{\bf e}_1\cdot\nabb\right]\vecn&&
{\bf e}_2\cdot\left[{\bf e}_1\cdot\nabb\right]\vecn\\ \\
{\bf e}_1\cdot\left[{\bf e}_2\cdot\nabb\right]\vecn
&&{\bf e}_2\cdot\left[{\bf e}_2\cdot\nabb\right]\vecn\end{matrix}\right]\end{equation}
The two dot products in the matrix are somewhat confusing.  To be concrete we write this matrix using its indices:
\begin{equation}\label{eindices}L_{ij} = -e^\alpha_i\ofss e^\beta_j\ofss {\partial n^\alpha\ofss\over\partial X^\beta\ofss}.\end{equation}
This matrix is also a tensor, known
as the {\sl Weingarten Map} or {\sl second fundamental form}\footnote{The metric tensor ${\bf g}$ is
sometimes called the {\sl first fundamental form}.}.  We can diagonalize this matrix via a similarity
transform ${\bm{\Lambda}}={\bf S}^{-1}{\bf L}{\bf S}$.  In this diagonal basis the entries of $\bm{\Lambda}$ are
precisely what we are after: the first entry in the upper left is the derivative of the surface normal
along a direction $\hat{\bf e}_1$.  Moreover, since the upper right entry vanishes, this derivative has
no components along $\hat{\bf e}_2$.  Therefore the upper left corner is the curvature of the curve
in the surface at $(s_1,s_2)$, tangent to $\hat{\bf e}_1$.  Similarly, the lower right entry of $\bm{\Lambda}$ gives
us another curvature.  Note that the two directions $\hat{\bf e}_1$ and $\hat{\bf e}_2$ remain orthogonal\footnote{The orthogonality of the two vectors $\hat{\bf e}_1$ and $\hat{\bf e}_2$ follows
from the fact the ${\bf S}^{-1}={\bf S}^T$. In other words, $\bf L$ can be diagonalized via an
orthogonal transformation since it is symmetric: $L_{12}-L_{21}=-\left(
e^\alpha_1 e^\beta_2 - e^\alpha_2 e^\beta_1\right)\partial_\beta n^\alpha =- {\bf n}\cdot\left[\nabb\times{\bf n}\right]=0$ when ${\bf n}$ is normal to 
a surface, {\sl i.e.} ${\bf n} = \nabb\phi/\vert\nabb\phi\vert$.}.  They
are known as the {\sl principal directions} on the surface and their associated curvatures $\kappa_1$ and $\kappa_2$ are
the {\sl principal curvatures} (see Figure 8). Equivalently, we can define the two principal radii of curvature through
$R_i=1/\kappa_i$.  We can easily extract these curvatures from the original matrix ${\bf L}$.  Note
that $\kappa_1\kappa_2=\det{\bf\Lambda}$ and $\kappa_1+\kappa_2=\Tr\;{\bf\Lambda}$.  Moreover, since the
trace is cyclic and $\det{\bf AB}=\det{\bf A}\det{\bf B}$, we see that:
\begin{eqnarray}\label{eidents}
\Tr\;{\bf L} &= \kappa_1\ofss +\kappa_2\ofss\nonumber\\
\det {\bf L} &= \kappa_1\ofss\;\kappa_2\ofss.\end{eqnarray}
The product of the curvatures is known as the {\sl Gaussian curvature}, $K=\kappa_1\kappa_2$, while the
average of the curvatures is the {\sl mean curvature}, $H={1\over 2}\left[\kappa_1+\kappa_2\right]$.  The Gaussian
and mean curvature contain all the information to describe the bending of our surface.
\begin{figure}\epsfxsize=3.5truein\epsfbox{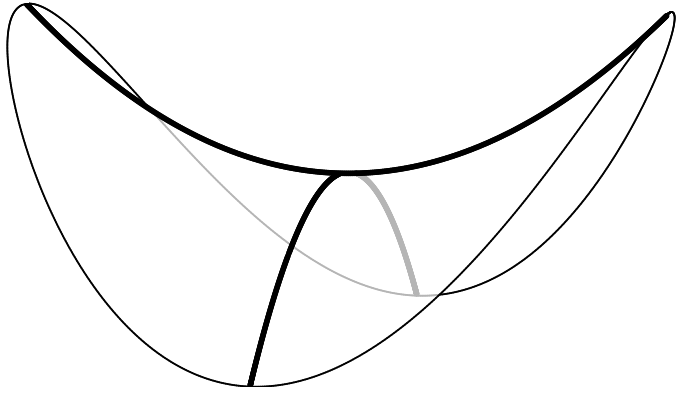}
\vskip10pt
\caption{Curves on a surface (heavy lines) with tangents along the principal directions
at the point of intersection.}
\end{figure}
\vskip10pt

The reader may be concerned that our definitions of curvature remain dependent on our choice of
basis vectors.  This is not the case.  Indeed, using the expression (\ref{eindices}) we have
\begin{eqnarray}\label{emeanandgauss}
H&=& -{1\over 2} \sum_{i=1}^2 e^\alpha_ie^\beta_i {\partial n^\alpha\over\partial x^\beta}\nonumber\\
K&=&\left(e_1^\alpha e_1^\beta e_2^\gamma e_2^\delta- e_1^\alpha e_2^\beta e_2^\gamma e_1^\delta \right)
{\partial n^\alpha\over\partial x^\beta}{\partial n^\gamma\over
\partial x^\delta}\end{eqnarray}
These expressions still look dependent on ${\bf e}_i$.  However, since $\vecn$ is a unit vector,
$n^\alpha n^\beta\partial_\beta n^\alpha=0$ and so we have
\begin{equation}\label{emeann}H = -{1\over 2} \left[e^\alpha_1e^\beta_1 + e^\alpha_2e^\beta_2 +n^\alpha n^\beta\right]{\partial n^\alpha\over
\partial x^\beta} = -{1\over 2}\delta^{\alpha\beta}{\partial n^\alpha\over\partial x^\beta}=-{1\over 2}\nabb\cdot\vecn,\end{equation}
where we have used the fact that $\{{\bf e}_1,{\bf e}_2,\vecn\}$ form an orthonormal triad.  Thus, as we hinted
at the beginning of this section, the mean curvature is proportional to the divergence of the surface normal.

We have to do a little more work on the expression for the Gaussian curvature.  The form in (\ref{emeanandgauss}) reminds
us of the expressions in the derivation of the Mermin-Ho relation.  Using the orthonormality of the basis vectors,
\begin{equation}\label{egaussn}K = e_1^\alpha e_2^\gamma \left( e_1^\beta e_2^\delta - e_2^\beta e_1^\delta\right)
\partial_\beta n^\alpha\partial_\delta n^\gamma =
e_1^\alpha e_2^\gamma \epsilon_{\beta\delta\rho}n^\rho \partial_\beta n^\alpha\partial_\delta n^\gamma.\end{equation}
Finally, as in the discussion of (\ref{ecurlofomegathree}), we have
\begin{equation}\label{egaussismh}K = {1\over2}\epsilon_{\alpha\beta\gamma}n^\gamma\epsilon_{ijk}n^k
\partial_i n^\alpha\partial_j n^\beta = {\bf n}\cdot\left[\nabb\cross{\bm{\Omega}}\right].\end{equation}
Moreover, the Gaussian curvature has a simple interpretation.   Consider the {\sl normal
spherical map} defined by analogy with the tangent spherical map.  For each point on the surface
we identify a point on the unit sphere which corresponds to the surface normal at that point.   This
map is also known as the {\sl Gauss map}.  Since
the surface element on our surface is  just $dS_\mu = n_\mu dS$, we see from the discussion at the end of \S{\bf III}C, that the Gaussian
curvature is just the area swept out by the surface normal on the Gauss map as we move
along the surface.  In other words, if we have a small region $M$ of our surface, then
\begin{equation}\label{elittlearea}\int_{M} K\,dS = \int_M {1\over2}\epsilon_{\alpha\beta\gamma}n^\gamma\epsilon_{ijk} 
\partial_i n^\alpha\partial_j n^\beta\,dS_k.\end{equation}
Thus the Gaussian curvature is the ratio between the infinitesimal area swept out on the Gauss map and
the infinitesimal area of the original surface to which it corresponds.

We have seen that the two principal curvatures depend only on the surface normal and not on our choice of
coordinates or basis vectors.  This is useful when we consider the energetics of fluid membranes.  They are
called ``fluid'' because they have no internal structure.  Therefore it would be unphysical to build an energy
out of anything but the two invariants $H$ and $K$.  Note that if we have an open surface there is no
distinction between inside and outside, so the layer normal $\vecn\ofss$ is defined only up to a sign.  Though
the Gaussian curvature is independent of this sign, the mean curvature is not.  To get around this, we
insist that the free energy be even in powers of $H$ \cite{canham,helfrich} :
\begin{equation}\label{efreesurf}F_{\rm CH}= \int dS \left\{2\kappa H^2 + \overline{\kappa} K\right\},\end{equation}
where $\kappa$ and $\overline{\kappa}$ are (confusingly) the standard symbols for the bending moduli.  This is
known as the {\sl Canham-Helfrich} free energy for fluid membranes. Note
that they both have units of energy, since the dimensions of $K$ and $H^2$ cancel the dimensions of the surface.  Also
note that the integration is done with respect to the actual surface area, so that $dS = \sqrt{\det{\bf g}}ds_1 ds_2$.
We will see in the following sections that the integral of the Gaussian curvature is a constant and so the
term proportional to $\overline{\kappa}$ is usually neglected.  We will also discover that not only is
the Gaussian curvature independent of our choice of basis vectors, but that it also can be measured
with no knowledge of the surface normal $\vecn\ofss$.  Finally, if the membrane is
{\sl tethered} then it has internal elastic degrees of freedom.  These degrees of
freedom couple to the geometry and produce a variety of singular structures \cite{didonna}.

\section{Global Theory of Surfaces}
\subsection{The Gauss-Bonnet Theorem: {\sl Foams on Curved Surfaces}}

\begin{figure}\epsfxsize=3.5truein\epsfbox{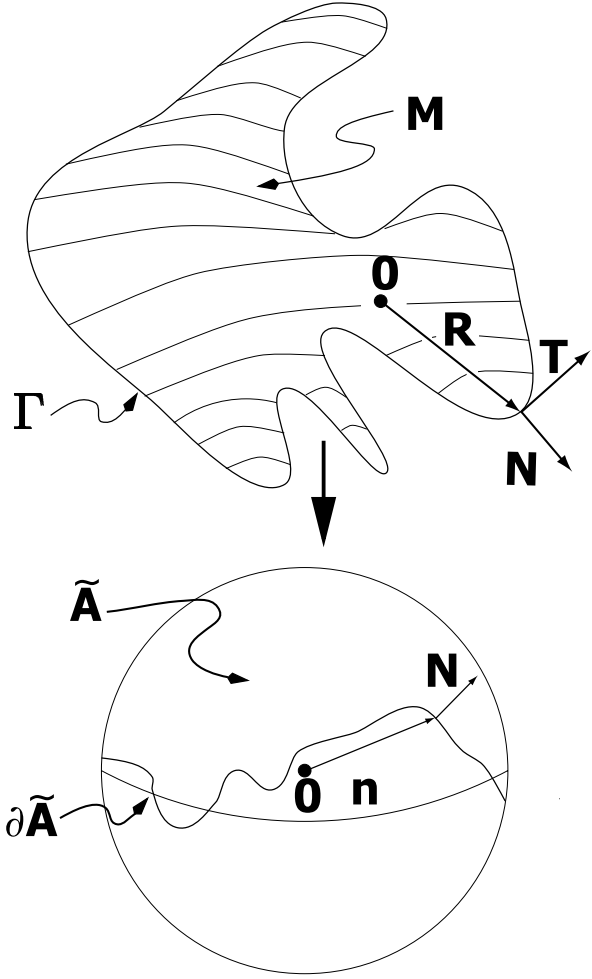}
\vskip10pt
\caption{Geometry for the Gauss-Bonnet Theorem.  Note that now we are
using the Gauss map which maps the surface normal ${\bf n}\ofx$ unto the unit sphere.}
\end{figure}
\vskip10pt

We have seen that there are two sorts of curvature that we can consider on
a surface, the mean and Gaussian curvatures.  We have studied both of these by considering
deviations of the normal vector to the surface.  However, a large part of differential
geometry focuses on {\sl intrinsic} properties, those quantities which
can be measured without reference to the space in which the manifold is embedded.
It turns out that the Gaussian curvature is intrinsic: by measuring the lengths and diameters of small circles entirely in the surface, one can determine $K$.

We have all the technology necessary to demonstrate this remarkable property.  To
show this, we consider a patch on our surface $M$, with boundary $\partial M$.  As in \S{\bf IV}A, we
can take the unit normal at each point and map it to the unit sphere.  As shown in Figure 9, when we traverse the
patch $M$ on the surface, we traverse a patch $\widetilde M$ on the unit sphere.  As we discussed,
the area of $\widetilde M$ is the integral of the Gaussian curvature of the patch $M$:
\begin{equation}\label{eNsurface}\widetilde A = \dint_{\widetilde M} d\widetilde S = \dint_M K dS = \dint_M {1\over 2}\epsilon_{\mu\nu\rho}\epsilon_{\alpha\beta\gamma}n^\alpha\ofx\partial_\nu
n^\beta\ofx\partial_\rho n^\gamma\ofx
dS_\mu\end{equation}
where we have used (\ref{egaussismh}).  Now consider the boundary curve $\partial M$.  This
curve has a tangent vector $\vecT(s)$ in space.   We know from our discussion of
curves that $\vecT'(s) = \kappa(s)\vecN(s)$.  However, while $\vecT(s)$ is also
tangent to the sphere, $\vecN(s)$ is, in general, {\sl not}.  The normal to the curve,
$\vecN(s)$, can have a component along the normal to the surface ${\bf n}(s)$ and perpendicular
to it.  If we lived in the surface and could only make measurements in the surface, like
some kind of bug, then we could only measure the component of the curve's normal
in the surface, the {\sl surface normal} ${\bm{\gamma}}(s)$:
\begin{equation}\label{esurfnorm}{\bm{\gamma}}(s) = {\vecT'(s) - {\left[\vecT'(s)\cdot{\bf n}(s)\right]}{\bf n}(s)\over
\kappa(s)\sqrt{1-\left[\vecN(s)\cdot{\bf n}(s)\right]^2}}\end{equation}
This is a unit vector which lies in the plane tangent to the surface and which is perpendicular to $\vecT(s)$, that is, the tangent to the boundary $\partial M$ of the patch $M$.  We would like to
define a curvature which measures the {\sl extra} curvature in the curve, not arising
from the curvature of the surface in which it is embedded.  By analogy to the
Frenet-Serret formulae, we define the {\sl geodesic curvature}\footnote{
We may also define the {\sl normal curvature}, $\kappa_n(s)$ which is the
curvature of our embedded curve that is imposed by the surface.  We define $\kappa_n(s)=\vecT'(s)\cdot{\bf n}(s)$ and thus the total curvature satisfies
$\kappa(s)=\sqrt{\kappa_n^2(s) + \kappa_g^2(s)}$.} $\kappa_g$ to be
\begin{equation}\label{egeocurv}\kappa_g(s)= \vecT'(s)\cdot{\bm{\gamma}}(s) = \kappa(s)\sqrt{1-\left[\vecN(s)\cdot
{\bf n}(s)\right]^2}.\end{equation}
Since $\{\vecT(s),{\bm{\gamma}}(s),{\bf n}(s)\}$ form an orthonormal triad, we have
\begin{equation}\label{egeocurvii}
\kappa_g(s) = \vecT'(s)\cdot\left[{\bf n}(s)\cross\vecT(s)\right] = {\bf n}(s)\cdot\left[\vecT(s)\cross\vecT'(s)\right],\end{equation}
a form we have seen before in the context of the Mermin-Ho relation!  Since
the unit tangent vector lies in the plane perpendicular to ${\bf n}\ofx$, we may write
it in terms of our arbitrary basis ${\bf e}_1\ofx$ and ${\bf e}_2\ofx$, introduced in
\S{\bf III}B, $\vecT(s)=\cos\theta{\bf e}_1 +\sin\theta{\bf e}_2$.  Using (\ref{egeocurvii}) to calculate
$\kappa_g$, we find that
\begin{equation}\label{ealmostthere}\kappa_g(s) = \partial_s\theta(s) - e_1^\alpha(s)\partial_s e_2^\alpha(s)\end{equation}
where we have used the fact that for a unit vector ${\bf u}(s)$, $\partial_s\left[{\bf u}(s)\right]^2=
2{\bf u}(s)\cdot\partial_s{\bf u}(s)=0$.  In analogy with  (\ref{eclosed}), we can integrate the
geodesic curvature around the boundary of $M$ to find:
\begin{equation}\label{earoundthebound}\oint_{\partial M} \kappa_g(s)ds = \oint_{\partial M}
\left[\nabb\theta\ofx - {\bm{\Omega}}\ofx\right] \cdot d{\bf R}.\end{equation}
As with our discussion of the relation between link, twist and writhe, we may transform the
line integral of ${\bm{\Omega}}\ofx$ into a surface integral over $M$ so that
\begin{equation}\label{egaussbo}\oint_{\partial M}\partial_s\theta(s)ds = \oint_{\partial M} \kappa_g(s)ds
+ \dint_M KdS\end{equation}
where we have used (\ref{emh}) to rewrite the final integrand as the Gaussian curvature (\ref{eNsurface}).
Finally, if the boundary curve does not intersect itself, then the tangent vector rotates around
${\bf n}\ofx$ exactly once, so $\theta(s)$ changes by $2\pi$ around the curve.  We have
thus established the {\sl Gauss-Bonnet Theorem}:
\begin{equation}\label{egb}\dint_M Kds + \oint_{\partial M}\kappa_g ds = 2\pi.\end{equation}
By integrating the geodesic curvature around a closed loop, we can calculate the
integrated Gaussian curvature that we surround.  As the loop shrinks ever smaller around
a point ${\bf x}_0$, we can calculate the Gaussian curvature at that point through division
by the surface area enclosed.  This is remarkable since the {\sl geodesic curvature can be measured without the use of the layer normal}!  In other words, since the geodesic curvature
is intrinsic, so is the Gaussian curvature!  We can take this result a step further by
considering regions that have discrete angles in their boundaries ({\sl i.e.} polygons).
When there is a sharp bend in the boundary curve $\partial M$, we cannot calculate
the geodesic curvature $\kappa_g(s)$.  We can, however, integrate the geodesic
curvature along the smooth parts of the boundary.  If there are $j$ sharp bends with
angles $\Delta\theta_j$, respectively, then when we integrate around the boundary
as in (\ref{earoundthebound}) we will have a deficit of $\sum_j\Delta\theta_j$, or in other
words the smooth part of $\theta(s)$ does not have to change by $2\pi$ for
the curve to come around.  We thus
have
\begin{equation}\label{egbtwo}\dint_M KdS +\oint_{\partial M}\kappa_g ds + \sum_j\Delta\theta_j = 2\pi\end{equation}
where we understand that the integral around the boundary should be broken
into smooth segments of the boundary.  The jump angles account for the
discontinuities.   Note that we can have lines for which $\kappa_g(s)=0$.  These
are the ``straight'' lines on the surface and are called {\sl geodesics}.  If we were
to build a polygon with $n$-sides, all of which are geodesics, then
(\ref{egbtwo}) reads:
\begin{equation}\label{edeficit}2\pi - \sum_j\Delta\theta_j = \dint_M KdS.\end{equation}
Since the sum of the external angles $\Delta\theta_j$ of a polygon
in flat space is $2\pi$, we see that the Gaussian curvature is a measure of
the excess (or deficit) angle in a polygon.  If the Gaussian curvature is positive the sum
of the external angles is smaller than $2\pi$ so the sum of the {\sl internal} angles
$\sum_j(\pi - \Delta\theta_j) = (n-2)\pi + \int\!\!\!\int\! KdS$ is larger than we might expect.  For
example, imagine the triangle on the sphere connecting the North Pole ($90^\circ$N), Pontianak, Indonesia (roughly $0^\circ$N, $109^\circ20'$W)
and Loolo, The Congo (roughly $0^\circ$S, $19^\circ20'$W) along great circles.  At each vertex
of this triangle the arcs of great circles meet at $90^\circ$ and so the sum of the
interior angles is $3\pi/2$.  Since $n=3$ for a triangle, we discover that $\int\!\!\!\int\! KdS = \pi/2$
for this triangle that covers ${1\over 8}$ of the globe.  However, we know that for
a sphere of radius $R$, the Gaussian curvature is just $1/R^2$ and so
$\int\!\!\!\int \!KdS = {1\over 8} {4\pi R^2\over R^2} = {\pi/2}$, as we expect!

There are many beautiful uses and examples of the interplay between Gaussian curvature
and geodesic curvature in physical systems.  For instance, Avron and Levine \cite{Levine} have
considered dry foams on curved, two-dimensional surfaces.  In this context, ``dry'' refers
to the fact that there is no fluid between adjacent bubbles so that the walls between
them can be treated as lines and the vertices may be treated as points.  There are two key ingredients to the physics
of dry foams: surface tension and pressure. In two dimensions the surface tension
amounts to a line tension $\sigma$ along the interfaces between neighboring bubbles.  If the
soap film is uniform then the line tension is constant as well.  Whenever the lines
meet, mechanical equilibrium must be maintained.   The pressure in the bubbles exerts
a force on their boundaries.  If two cells are separated by a boundary line then
the pressure difference $\Delta P$ must be balanced by the boundary.  As with the derivation of
the wave equation, the force exerted by the boundary curve $\Gamma$ is
$\sigma\vecT'(s) $.  However, the pressure exerts forces only within the surface
and so we are only interested in the component of the force in the plane of the surface.
According to the above discussion, that force is $\sigma\kappa_g(s){\bm{\gamma}}(s)$,
and so the magnitude of the force is $\sigma\kappa_g(s) = \Delta P$.  In the case
of planar foams, $\kappa_g(s)=\kappa(s)$ and this is
known as the Young-Laplace law.  To model the diffusion of gas from one cell
to the other, we assume a simple dynamics where
\begin{equation}\label{edynsoap}{dN\over dt} = - C\sum_j \Delta P_j \ell_j\end{equation}
where $N$ is the number of gas molecules in the cell of interest, $\Delta P_j$ is the pressure difference between it and
its $j^{\rm th}$ neighbor, $\ell_j$ is the length of the boundary separating the cell from
its $j^{\rm th}$ neighbor, and $C>0$ is a diffusion constant.  This dynamics captures the
simple idea that if a bubble is higher pressure than its neighbors so that $\Delta P_j>0$, then
it loses gas, while if it has lower pressure it gains gas.  Using the generalization of the
Young-Laplace law and the Gauss-Bonnet theorem we have:
\begin{equation}\label{ylgb}{dN\over dt} = - C\sigma\oint_{\partial M} \kappa_g(s) ds=
C\sigma\left\{\dint_M KdS  + \sum_j(\pi - \alpha_j) - 2\pi\right\}\end{equation}
where $\alpha_j$ are the internal angles.

For flat surfaces with $K=0$ it is
known that a hexagonal honeycomb network of boundaries minimizes the
length of the cell walls \cite{Hales,Morgan}.  If we imagine starting with a stable configuration
on a flat membrane and distorting the membrane, we should take each internal angle to
be $2\pi/3$.  Doing so Avron and Levine \cite{Levine} found:
\begin{equation}\label{eal}{dN\over dt} = C\sigma\left\{ {\pi\over 3}(n-6) + \dint_M KdS \right\}.\end{equation}
If we seek a stationary configuration so that $N$ is time independent, we
see that for flat membranes $K=0$ and only bubbles with $n=6$ sides are stationary: hexagons.
If the surface has positive curvature, $K>0$, and there is an instability.  To make $(dN/dt)=0$, we
see that $n<6$.  However, even if an area and an $n$ were found to make $N$ time
independent, we can see that there is an instability:  if the area grows then $(dN/dt)$ becomes
positive so the bubble grows {\sl more}.  Likewise, if the bubble shrinks then the integral over $K$
gets smaller and so gas flows out of the bubble and it shrinks some more.  Thus on a
positively curved surface the only stable situation is one for which one bubble overtakes
the whole system.
On the other hand, when the surface has negative Gaussian curvature,  it follows from a similar
argument that when the area of the bubble increases $(dN/dt)<0$ and when the area shrinks
$(dN/dt)>0$.  In the critical case of a flat membrane the stationary solution is a hexagon
and it is neither stable nor unstable towards growth \cite{Levine}.

\subsection{The Euler Characteristic and the Genus: {\sl Defects on Surfaces}}

We can take (\ref{egbtwo}) one step further by considering a closed surface.  In this
case we can integrate the Gaussian curvature over the whole manifold.  If we triangulate
the entire surface to form a net, then we can use the Gauss-Bonnet theorem to establish
a relation between the topology of the network and the total Gaussian curvature.   At
each vertex there is a total angle of $2\pi$ which is divided into the internal angles
of the triangles meeting at that point.  Each face contributes $\pi$ to the total angle
at the vertices in addition to the excess angle from that face, $\int\!\!\!\int\!KdS$.  Adding all
the triangles together we have:
\begin{equation}\label{eeuler}2\pi V = \pi F + \doint_M KdS\end{equation}
where $V$ is the number of vertices and $F$ is the number of faces.  Each face contributes
three edges $E$, but each edge is shared by two triangles, so $3F=2E$.
We then have
\begin{equation}\label{eeulerii}
V-E +F = {1\over 2\pi} \doint_M KdS.\end{equation}
This is a remarkable result.  As we show in the bottom two graphs of Figure 10 , if we remove an edge from our network
($E\rightarrow E-1$), two
faces join into one ($F\rightarrow F-1$), and we lose two vertices ($V\rightarrow V-2$), so
$V-E+F$ does not change!  Likewise if we add an edge $V-E+F$ is unchanged.  Thus
our result does not require the use of triangles.  This invariant is known as
the {\sl Euler characteristic} $\chi=V-E+F$.  Note that if we have any network that can be deformed into
a sphere
without cutting the edges or changing the vertices, then $\chi$ can be calculated using
(\ref{eeulerii})  for a perfect sphere, where the surface area is $4\pi R^2$ and the Gaussian
curvature is $1/R^2$.  Thus we find that
\begin{equation}\label{eeuleriii}V-E+F=\chi =2\end{equation}
for a network with the topology of a sphere.  Suppose we take two faces on the
sphere, deform them to be the same triangle and place them together, face-to-face.  In
the process we lose three edges ($E\rightarrow E-3$), two faces ($F\rightarrow F-2$) and
three vertices ($V\rightarrow V-3$) so $\chi=0$.  What have we done?  We have made
a doughnut (or a torus) with one handle and have reduced the
Euler characteristic by two.  Clearly, any time we add a handle $\chi$ is reduced by two.  If we
define the {\sl genus} $g$ to be the number of handles of the surface, then
\begin{equation}\label{egenus}{1\over 2\pi}\doint_M KdS = V- E +F =\chi = 2(1-g).\end{equation}
Though the Gaussian curvature was a geometric property, when integrated
over the entire surface it becomes a topological invariant, independent of the
local geometry.
\vskip10pt
\begin{figure}\epsfxsize=5.5truein
\epsfbox{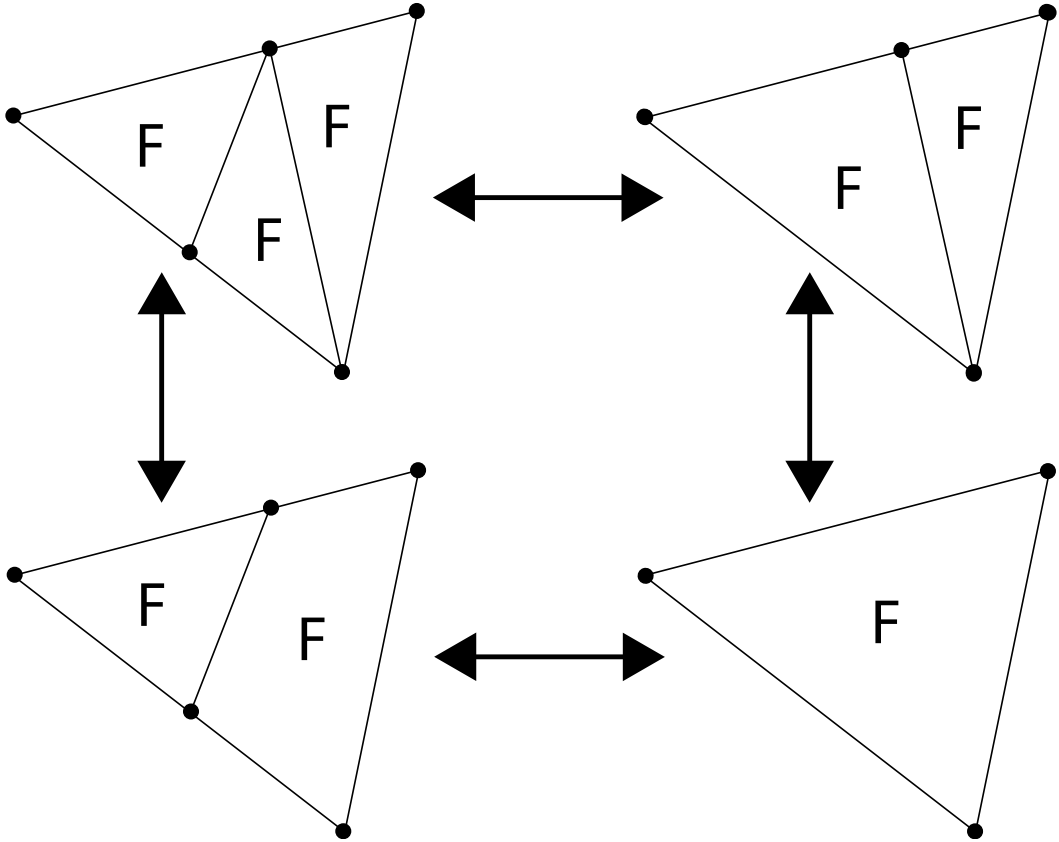}
\vskip10pt
\caption{Removing or adding an edge from a graph maintains the value of the
Euler characteristic
$V-E+F$.}
\end{figure}
\vskip10pt

The Euler characteristic can be used to understand the topology of defects on closed
surfaces.  Suppose that we have a unit vector field ${\bf v}\ofx$ living in the local
tangent plane to a closed
surface of genus $g$.    Since ${\bf v}\ofx$ lies in the tangent plane, we have ${\bf v}\ofx = \cos\theta\ofx {\bf e}_1\ofx + \sin\theta\ofx
{\bf e}_2\ofx$, where we have reintroduced our vectors ${\bf e}_1\ofx$ and ${\bf e}_2\ofx$
which are everywhere perpendicular to the unit normal ${\bf n}\ofx$ of the surface.
We will try to cover the surface with a
vector field that is single valued.  In other words, if we integrate derivatives of ${\bf v}\ofx$
around a closed curve, we should get back the same vector.
But as we discussed in \S{\bf III}B, to ensure this property we should focus
on the covariant derivative ${\bf D}\theta\ofx$.  We will have a single-valued
vector field when the curl of this derivative vanishes.
In this case equations (\ref{emh}) and (\ref{egaussismh}) give:
\begin{equation}\label{etotaldefect}
\doint_M \left[\nabb\cross\nabb\theta\ofx\right]\cdot d{\bf S} = \doint_M KdS = 2\pi\chi\end{equation}
Usually $\nabb\cross\nabb\theta=0$, but this result tells us that it is not
true on an arbitrary genus surface $M$.

\begin{figure}\epsfxsize=2.5truein\epsfbox{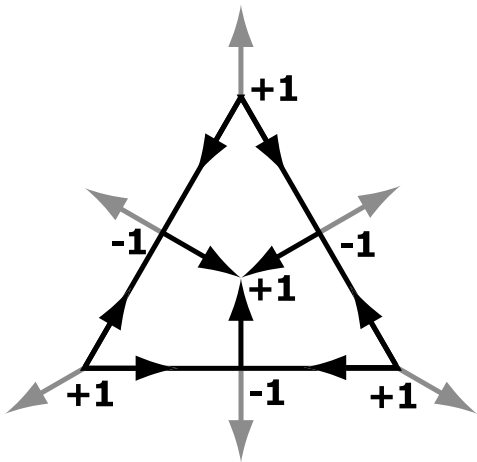}
\vskip10pt
\caption{A vector field on a triangular patch.  Note that the $+1$ defects
at the vertices and at the center force $-1$ defects on the edges.}
\end{figure}
\vskip10pt

Without presenting a detailed discussion on topological defects \cite{cl}, we can
see that $\nabb\cross\nabb\theta\ofx$ does not vanish in the presence of
a defect configuration and that on a two dimensional surface:
\begin{equation}\label{evortex}\nabb\cross\nabb\theta\ofx = \sum_i m_i \delta^2({\bf x} - {\bf x}_i),\end{equation}
where $m_i$ is the charge of the defect and ${\bf x}_i$ is the corresponding
position.  To see this, we consider
\begin{equation}\label{evorr}\theta\ofx = m\tan^{-1}\left({y\over x}\right).\end{equation}
It is straightforward to calculate $\nabb\theta\ofx$ and then to integrate the gradient
around a closed curve $\gamma$ that contains the origin.  We have
\begin{eqnarray}\label{eintegralvor}
2\pi m &=&\oint_\gamma \nabb\theta\ofx \cdot d{\bf x} \nonumber\\
&=&\dint_M \nabb\cross\nabb\theta\ofx\cdot d{\bf S},\end{eqnarray}
where $\gamma$ is the boundary of $M$.  Since we may arbitrarily
shrink the path $\gamma$ around the origin and the integral remains
constant, it must be that $\nabb\cross\nabb\theta = m\delta^2({\bf x})$.  Generalizing
this to an arbitrary collection of defects leads us to (\ref{evortex}).  Using (\ref{evortex}) in
(\ref{etotaldefect}) gives us:
\begin{equation}\label{eindexthm} \sum_i m_i = 2(1-g)=\chi.\end{equation}
This result is known as the {\sl Poincar\'e-Brouwer theorem}.  Since a vector
field must be single valued on the surface, $m_i$ must be an integer so that
$\cos\theta$ and $\sin\theta$ are well-defined.  This result tells us, for instance, that
a vector field on the surface of a sphere must have two $+1$ defects or one $+2$ defect.  On
a torus, however, no defects are necessary.

\begin{figure}\epsfxsize=3.5truein\epsfbox{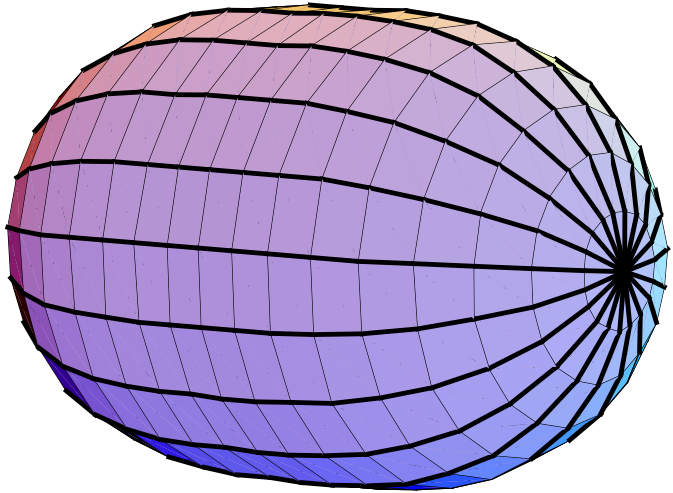}
\vskip10pt
\caption{A vesicle with some sort of vector order parameter.  The Poincar\'e-Brouwer theorem assures that there are two defects in the vector field, drawn as the dark lines
on the vesicle.  The vesicle distorts so that the Gaussian curvature is larger in the vicinity
of the defects.}
\end{figure}
\vskip10pt

There is another, more geometric way to establish (\ref{eindexthm}).  We first argue
that two vector fields on the same surface must have the same total topological
charge.  Consider two vector fields ${\bf u}\ofx$ and ${\bf v}\ofx$.  We can triangulate
the surface so that each triangle contains at most one defect in ${\bf u}\ofx$ and
one defect of ${\bf v}\ofx$ and that every defect is in some triangle.  In each
triangle we may now integrate $\nabb\theta_u\ofx$ and $\nabb\theta_v\ofx$ around
the triangle boundary.  The total topological charge $\chi_u$ of ${\bf u}\ofx$ is
\begin{equation}\label{etotalcharge}\chi_u = \sum_{\rm triangles}\,\sum_{\rm edges}\int\nabb\theta_u\ofx\cdot d{\bf x}\end{equation}
and similarly for ${\bf v}\ofx$.  Note that to establish the angles $\theta_u\ofx$ and
$\theta_v\ofx$, we must choose basis vectors ${\bf e}_i\ofx$. Since we know that
$\nabb\theta_u\ofx$ depends on our choice of these vectors, the integrals
of $\nabb\theta_u\ofx$ around each triangle depend on the underlying geometry
through the Mermin-Ho relation (\ref{emh}).  More importantly, each
vector field on $M$ may require a different choice of basis vectors ${\bf e}_i\ofx$, since
a defect in any particular triangle forces a constraint on ${\bf e}_i\ofx$.  Thus it
is not, in general, possible to calculate $\chi_u$ and $\chi_v$ using the same set
of ${\bf e}_i\ofx$.  However, the
difference between $\chi_u$ and $\chi_v$ is
just
\begin{equation}\label{edifference}\chi_u-\chi_v = \sum_{\rm triangles}\,\sum_{\rm edges}\int\nabla\left[\theta_u\ofx - \theta_v\ofx\right] \cdot d{\bf x} = 0.\end{equation}
The difference vanishes since the angle between ${\bf u}\ofx$ and ${\bf v}\ofx$ is
independent of our choice of basis vectors.  Therefore when we sum over all
triangles we get each edge twice but in opposite directions.  Since we
do not need the basis vectors to define the integrals, it is clear that the integrals
in opposite directions cancel and we get $0$.   Thus any two vector fields have
the same topological charge.

We will now count the defects of a vector field that we construct on triangulated surface, via the
following rules:
(1) put a defect with $+1$ charge at each vertex,
(2) put a defect with $+1$ charge at each center, and
(3) put a defect with $-1$ charge on each edge.
We can see that this will produce a consistent vector field everywhere else, as shown in Figure 11.  Adding
together the charges we have
\begin{equation}\label{eaddagain}\chi_u=V-E+F=\chi,\end{equation}
and so the topological charge is the Euler characteristic for any vector field on the surface.

A striking example of the effect of these defects can be seen when there is
a coupling between the bending of the surface and the configuration of the vector
field \cite{maclub,maclubii} .   For instance, if we have a vector field living on a closed surface then
it must have two $+1$ defects.  But because the defects are discrete, when we consider
(\ref{etotaldefect}) we find that the Gaussian curvature $K$ must also be confined to
the core of the defect since we may also integrate (\ref{emh}) over any submanifold of $M$.  But
then all the Gaussian curvature sits at the defects.  If instead of viewing (\ref{emh}) as a constraint,
we write it as an energy:
\begin{equation}\label{efreeforsurf}F = {1\over 2}\int dS\, A\left[\nabb\theta\ofx - {\bm{\Omega}}\ofx\right]^2 + F_{\rm CH},\end{equation}
where $F_{\rm CH}$ is the free energy for the surface (\ref{efreesurf}), and $A$ is the spin-stiffness.
On a flat surface we can always choose ${\bm{\Omega}}={\bf0}$ and so the first term in
(\ref{efreeforsurf}) just describes the usual Goldstone mode of broken rotational invariance. On a curved
surface the inclusion of ${\bm{\Omega}}$ is necessary in order to make the free energy
uniquely defined.  The energy (\ref{efreeforsurf}) now represents a competition between
surface bending and the requirement of two $+1$ topological defects which serve as 
a source for Gaussian curvature.  Both of these energies vanish for the plane, but
if we restrict our topology to a sphere, for instance, then neither term can be made
to vanish everywhere.  Balancing these two effects, a vesicle would form an oblate
shape, as shown in Figure 12.

\section{Three Dimensions and Beyond}
We have covered the basic elements of the geometry of curves and
surfaces.  Things become significantly more abstract in three dimensions because
there is no analog of the normal vector, binormal vector or surface normal: the
three dimensional system is not embedded in a higher dimensional space.  Though it
might be hard to visualize how something can have curvature without an
embedding space, we already know that it is possible.  Recall that the Gaussian
curvature could be measured without any knowledge of the surface normal.  Quantities
with that property are called {\sl intrinsic}.   The general theory of relativity relies
on intrinsic quantities to describe the curvature of four-dimensional space-time \cite{MTW} --
there is no reference to a larger space in which our universe lives.
Though we will not go into any mathematical detail, it is worthwhile to describe
two examples from soft materials.

The first is the blue phases of chiral liquid crystals \cite{meiboom,Wright}.  These are phases
in which there is a three-dimensional, periodic modulation of the nematic director ${\bf n}\ofx$ with
a lengthscale comparable to that of visible light.  These usually exist only over
a narrow temperature range and are stabilized by an often neglected term in the
nematic free energy.  The Frank free energy for a nematic liquid crystal is
\begin{eqnarray}\label{nematic}
F &=& \frac{1}{2}\int d^3\!x\,\Big\{K_1\left(\nabb\cdot{\bf n}\right)^2 + K_2\left[{\bf n}\cdot\left(\nabb\cross{\bf n}\right)\right]^2 + 
K_3\left[{\bf n}\cross\left(\nabb\cross{\bf n}\right)\right]^2 \nonumber\\ &&\qquad+2K_{24}\nabb\cdot\left[\left({\bf n}\cdot\nabb\right){\bf n} - \left(\nabb\cdot{\bf n}\right){\bf n}\right]\Big\}
\end{eqnarray}
The standard elastic constants $K_1$, $K_2$ and $K_3$ are a measure of the
energy cost for splay, twist and bend modes, respectively \cite{cl}.  The last term with
elastic constant $K_{24}$ is known as the {\sl saddle-splay} term.  Though it is a total
derivative, when there are defects present it can contribute to the energy.  The blue
phase is riddled with precisely those defects that contribute to saddle-splay. The remarkable
thing about this term is that it is precisely the Gaussian curvature as in (\ref{egaussismh}):
\begin{equation}\label{ssisg}
{\bf n}\ofx\cdot\nabb\cross\bm{\Omega}\ofx = - \frac{1}{2}\nabb\cdot\left[\left({\bf n}\cdot\nabb\right){\bf n} - \left(\nabb\cdot{\bf n}\right){\bf n}\right]
\end{equation}
What does this mean?  If there are surfaces to which the nematic director is normal, then
the saddle-splay is the Gaussian curvature of those surfaces.  However, the
saddle-splay is more general.  Even if the director is not a field of layer normals, the
saddle-splay is a measure of curvature.  Indeed, the blue phases can be understood as
unfrustrated systems in curved three-dimensional space \cite{sethna}.  Projecting this texture
into flat space leads to the topological defects in these phases.  A compelling and alternative way
of viewing the blue phases is to view them as decorations of space filling minimal surfaces \cite{pansu}.  It is the connection between topological defects and curvature that
makes it possible to identify and locate defects in lattice simulations of liquid crystals \cite{pelcovits}.

Another three-dimensional system that can be understood in terms of curvature is 
the melting and freezing of hard-sphere fluids.  In two dimensions,  as hard-spheres condense from the fluid phase to the crystalline phase, they form close-packed triangles.  Eventually these can assemble into a triangular lattice.  The situation is not so happy in three-dimensions.  There, four spheres can close-pack into a tetrahedron, but tetrahedra cannot assemble to fill space.  As a result, all lattices have a packing density lower than the best local packing.  This difficulty can be viewed as a geometric frustration along the same lines as the blue phase.  In positively curved space, however, this frustration can also be eliminated \cite{nelsong,widom}, and tetrahedral close-packing can fill space.  The BCC phase
is the simplest example of this structure in flat space.

\section*{Outlook}
Though the mathematics
described here has a certain elegance and beauty,
I hope that I have conveyed the utility of differential geometry in a variety of
physical problems.  While a geometric description of a system often leads to an intuitive
perspective, there are many other arenas in which a geometric
formulation of the problem is not only useful but essential.  Put together with
statistical mechanics, differential geometry has been and will continue to be a powerful
tool in the study of soft materials.

\begin{acknowledgements}
I thank I. Bluestein, B.A. DiDonna, G.M. Grason, M.I. Kamien, W.Y. Kung, T.C. Lubensky, T.R. Powers, P. Ziherl and especially G. Jungman and P. Nelson for careful and critical reading of this review. This work was supported by NSF Grants DMR01-29804 and INT99-10017,
and a gift from L.J. Bernstein.
\end{acknowledgements}
\newpage


\end{document}